\documentclass[useAMS,usenatbib]{mn2e}
\usepackage[dvips]{graphicx}
\usepackage{hyperref}
\usepackage{xcolor}
\usepackage{caption}
\newcommand{\bz}{$\langle B_z \rangle$}
\newcommand{\nz}{$\langle N_z \rangle$}
\newcommand{\vsini}{$v \sin i$}
\newcommand{\kms}{km\,s$^{-1}$}

\newcommand{\mdot}{$\dot{M}$}
\newcommand{\vinf}{$v_\infty$}

\newcommand{\msun}{M$_\odot$}

\newcommand{\teff}{$T_{\rm eff}$}
\newcommand{\ra}{$R_{\rm A}$}
\newcommand{\rk}{$R_{\rm K}$}

\newcommand{\mnras}{$MNRAS$}
\newcommand{\apj}{$ApJ$}
\newcommand{\apjl}{$ApJL$}
\newcommand{\aj}{$AJ$}
\newcommand{\aap}{A\&A}

\newcommand{\aaps}{A\&AS}

\newcommand{\apjs}{ApJS}
\newcommand{\araa}{ARAA}

\newcommand{\nat}{Nature}
\newcommand{\procspie}{$SPIE$}
\newcommand{\physscr}{$Physica Scripta$}

\title[HD 156424]{MOBSTER -- V: Discovery of a magnetic companion star to the magnetic $\beta$ Cep pulsator HD 156424\thanks{Based on observations obtained at the Canada-France-Hawaii Telescope (CFHT) which is operated by the National Research Council of Canada, the Institut National des Sciences de l'Univers of the Centre National de la Recherche Scientifique of France, and the University of Hawaii; and at the La Silla Observatory, ESO Chile with the MPA 2.2 m telescope.}}
\author[M. E. Shultz]
{M.\ E.\ Shultz$^{1}$\thanks{E-mail: mshultz@udel.edu},
Th.\ Rivinius$^{2}$,
G.\ A.\ Wade$^{3}$,
O.\ Kochukhov$^{4}$,
E.\ Alecian$^{5}$,
\newauthor{A. David-Uraz$^1$, J.\ Sikora$^6$, and the MiMeS Collaboration} \\
$^1$Department of Physics and Astronomy, University of Delaware, 217 Sharp Lab, Newark, Delaware, 19716, USA\\
$^2$ESO - European Organisation for Astronomical Research in the Southern Hemisphere, Casilla 19001, Santiago 19, Chile\\
$^3$Department of Physics, Royal Military College of Canada, Kingston, Ontario K7K 7B4, Canada\\
$^4$Department of Physics and Astronomy, Uppsala University, Box 516, Uppsala 75120 \\
$^5$Univ. Grenoble Alpes, CNRS, IPAG, 38000 Grenoble, France\\
$^6$Department of Physics and Astronomy, Bishop's University, Sherbrooke, Qu{\'e}bec, Canada, J1M 1Z7\\
}
\begin{document}

\date{}

\pagerange{\pageref{firstpage}--\pageref{lastpage}} \pubyear{2002}

\maketitle

\label{firstpage}

\begin{abstract}
HD\,156424 (B2\,V) is a little-studied magnetic hot star in the Sco OB4 association, previously noted to display both high-frequency radial velocity (RV) variability and magnetospheric H$\alpha$ emission. We have analysed the TESS light curve, and find that it is a $\beta$ Cep pulsator with 11 detectable frequencies, 4 of which are independent $p$-modes. The strongest frequency is also detectable in RVs from ground-based high-resolution spectroscopy. RVs also show a long-term variation, suggestive of orbital motion with a period of $\sim$years; significant differences in the frequencies determined from TESS and RV datasets are consistent with a light-time effect from orbital motion. Close examination of the star's spectrum reveals the presence of a spectroscopic companion, however as its RV is not variable it cannot be responsible for the orbital motion and we therefore infer that the system is a hierarchical triple with a so-far undetected third star. Reanalysis of LSD profiles from ESPaDOnS and HARPSpol spectropolarimetry reveals the surprising presence of a strong magnetic field in the companion star, with \bz~about $+1.5$~kG as compared to \bz~$\sim -0.8$~kG for the primary. HD\,156424 is thus the second hot binary with two magnetic stars. We are unable to identify a rotational period for HD\,156424A. The magnetospheric H$\alpha$ emission appears to originate around HD\,156424B. Using H$\alpha$, as well as other variable spectral lines, we determine a period of about 0.52~d, making HD\,156424B one of the most rapidly rotating magnetic hot stars.
\end{abstract}

\begin{keywords}
stars: individual: HD 156424 -- stars: binaries: spectroscopic -- stars: early-type -- stars: magnetic field -- stars: oscillations
\end{keywords}

\section{Introduction}

   \begin{figure*}
   \centering
   \includegraphics[width=18cm]{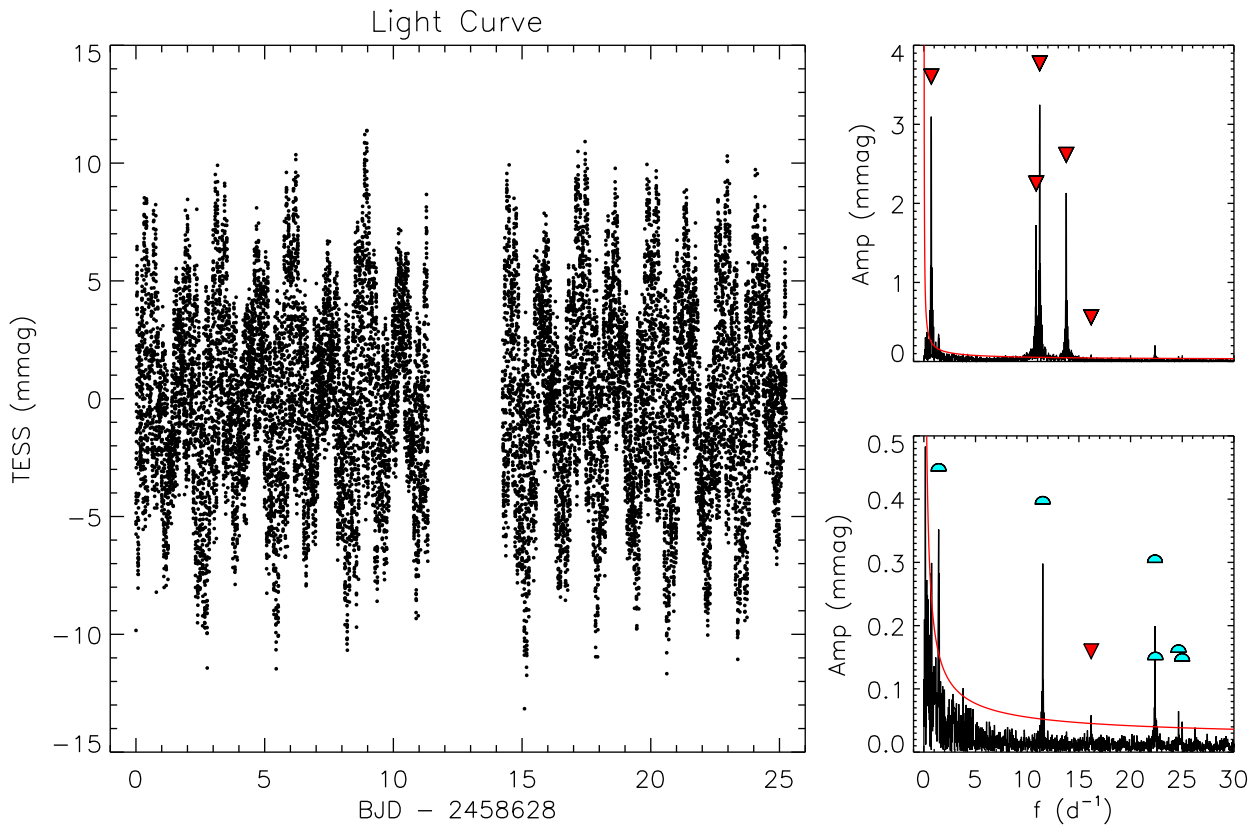}
      \caption[]{{\em Left panel}: TESS light curve. {\em Right panels}: Full TESS light curve frequency spectrum ({\em top}) and the frequency spectrum after pre-whitening with the 4 strongest frequencies ({\em bottom}). The red line shows $S/N = 4$. Primary frequencies are indicated with red triangles, combinations and harmonics with blue half-circles.}
         \label{hd156424_tess_lc}
   \end{figure*}

Approximately 10\% of early-type stars are magnetic \citep[e.g.][]{2017MNRAS.465.2432G,2019MNRAS.483.2300S}. In contrast to the magnetic fields of stars with convective envelopes, which are sustained by contemporaneous rotational-convective dynamos \citep[e.g.][]{2008MNRAS.390..545D,2016MNRAS.457..580F,2018MNRAS.474.4956F}, no dynamo mechanism has been confirmed to be sustainable in the radiative envelopes of hot stars. This has led to the suggestion that massive star magnetic fields are `fossils', remnants of a previous era in the star's formation which retain their stability in highly conductive radiative envelopes \citep[e.g.][]{2004Natur.431..819B,2009MNRAS.397..763B,2010ApJ...724L..34D,2015IAUS..305...61N}. A fossil origin is consistent with the observed properties of hot star magnetic fields: they are topologically simple \citep[typically `distorted dipoles';][]{2019A&A...621A..47K}; stable over at least decades \citep[e.g.][]{2018MNRAS.475.5144S}; their unsigned magnetic flux is either conserved \citep{2019MNRAS.483.3127S} or slowly decays \citep{land2007,land2008,2016A&A...592A..84F,2019MNRAS.490..274S} over evolutionary timescales; and there is no clear correlation between the surface magnetic field strength and physical properties such as rotation \citep[e.g.][]{2019MNRAS.483.3127S,2019MNRAS.490..274S}, in contrast to what is unambiguously observed for stars with convective envelopes \citep[e.g.][]{2016MNRAS.457..580F,2018MNRAS.474.4956F}.

The external properties of magnetic hot stars are relatively well understood, but little is known about the internal magnetic configurations of stars with fossil fields, resulting in great uncertainty regarding evolutionary models incorporating fossil magnetism \citep[][]{2019MNRAS.485.5843K,2020MNRAS.493..518K,2020svos.conf..293T}. There is some evidence that extremely strong magnetic fields can suppress convection in opacity-bump convection zones, and thereby reduce or eliminate macroturbulence \citep{2013MNRAS.433.2497S}. Asteroseismological analysis has suggested that core overshooting may be suppressed by fossil magnetic fields \citep{2012MNRAS.427..483B}. Since the best method of probing stellar interiors is via asteroseismology, identification of appropriate magneto-asteroseismic targets is a high priority \citep[e.g.][]{2018A&A...616A..77B,2018MNRAS.478.2777B,2019A&A...622A..67B}. Efforts to expand the sample of asteroseismic targets on the upper main sequence are an important part of this, since about 10\% of them can be expected to be magnetic \citep[e.g.][]{2019ApJ...872L...9P,2019MNRAS.489.1304B,2020A&A...639A..81B,2020AJ....160...32L}.

HD\,156424 is a B2\,V star in the Sco OB4 association. \cite{alecian2014} detected the star's magnetic field, finding a longitudinal magnetic field of about $-500$~G. Further observations were analyzed by \cite{2018MNRAS.475.5144S}, who confirmed the low level of variation in the magnetic field and reported a tentative periodicity of about 2.8~d. Radial velocity (RV) variation of a few \kms~was reported by \cite{alecian2014}.

\cite{alecian2014} also reported the presence of H$\alpha$ emission consistent with an origin in a Centrifugal Magnetosphere \citep[CM; e.g.][]{lb1978,town2005c,petit2013}. In most cases, CM-type H$\alpha$ emission occurs above the Kepler corotation radius \rk, which for the rapidly rotating, strongly magnetized stars that host such emission is generally at a distance of a few stellar radii \citep[e.g.][]{2019MNRAS.490..274S}, or equivalently a few times \vsini. HD\,156424's H$\alpha$ emission is anomalous in this regard because it peaks at about 20 to 30 times \vsini, apparently much larger than the star's Alfv\'en radius \citep{2019MNRAS.490..274S}, i.e.\ at a greater distance than the maximum extent of magnetic confinement \citep{ud2002}. HD\,156424's magnetosphere has also been detected in gyrosynchrotron emission \citep{2017MNRAS.465.2160K}, although not yet in X-rays \citep{2014ApJS..215...10N}.

The rapid radial velocity variation opens the possibilty that HD\,156424 may be one of the rare class of pulsating magnetic stars. Determining if this is the case, and if so providing an initial characterization of its pulsation properties, provided the initial motivation for our analysis of the recently obtained Transiting Exoplanet Survery Satellite \citep[TESS;][]{2015JATIS...1a4003R} light curve together with the existing ground-based spectroscopic dataset. Further motivation was provided by the anomalous nature of the H$\alpha$ emission noted by \cite{2020arXiv200912336S}, and the ambiguous rotation period inferred from magnetic data by \cite{2018MNRAS.475.5144S}. A description of the available datasets is provided in \S~\ref{sec:obs}, while the frequency analyses of the light curve and RVs are respectively described in \S~\ref{sec:lc} and \S \ref{sec:rv}. The RV analysis provided unexpected evidence that the star has a companion, in consequence of which we analyse the spectra looking for evidence of binarity in \S \ref{sec:multiplicity}, following which we determine the atmospheric properties of the components. A reanalysis of the magnetic and H$\alpha$ data in the light of multiplicity is provided in \S~\ref{sec:mag} and \S \ref{sec:halpha}, with the surprising results that both stars are magnetic, and that the H$\alpha$ emission almost certainly originates around the secondary. Magnetic models are inferred in \S~\ref{sec:magpars}. The conclusions are summarized in \S~\ref{sec:discussion}.

\section{Observations}\label{sec:obs}

\subsection{ESPaDOnS spectropolarimetry}

ESPaDOnS is a fibre-fed echelle spectropolarimeter mounted at the Canada-France-Hawaii Telescope (CFHT). It has a spectral resolution $\lambda/\Delta\lambda \sim 65,000$, and a spectral range from 3700 to 10500 \AA~over 40 spectral orders. Each observation consists of 4 polarimetric sub-exposures, between which the orientation of the instrument's Fresnel rhombs are changed, yielding 4 intensity (Stokes $I$) spectra, 1 circularly polarized (Stokes $V$) spectrum, and 2 null polarization ($N$) spectra, the latter obtained in such a way as to cancel out the intrinsic polarization of the source. \cite{2016MNRAS.456....2W} describe the reduction and analysis of ESPaDOnS data in detail. Nine Stokes $V$ observations were acquired between 04/2014 and 06/2014 by a P.I. program\footnote{Program Code CFHT 14AC010}. A uniform sub-exposure time of 450 s was used for all observations. The median peak signal-to-noise ratio ($S/N$) per spectral 1.8~\kms~pixel is 369. 

\subsection{HARPSpol spectropolarimetry}

HARPSpol is a high-resolution ($\lambda/\Delta\lambda~\sim~110,000$) echelle spectropolarimeter with a spectral range covering 3780--6910 \AA, with a gap between 5240 and 5360 \AA, across 71 spectral orders. It is installed at the 3.6~m telescope at the European Southern Observatory (ESO) La Silla facility. As with ESPaDOnS, each spectropolarimetric sequence consists of 4 polarized sub-exposures, which are combined to yield the Stokes $V$ spectrum as well as a diagnostic null $N$. The sub-exposure time was 900~s. Three observations were acquired in 2012 by the Magnetism in Massive Stars (MiMeS) ESO Large Program. The acquisition, reduction, analysis, and characteristics of these data was described by \cite{alecian2014}.

\subsection{FEROS spectroscopy}

FEROS is a high-dispersion echelle spectrograph, with $\lambda/\Delta\lambda\sim 48,000$ and a spectral range of $3750-8900$~\AA \citep{1998SPIE.3355..844K}. It is mounted at the 2.2~m La Silla MPG telescope. We acquired 11 spectra between 06/2015 and 07/2015, with an exposure time of 1400 s. The data were reduced using the standard FEROS Data Reduction System MIDAS scripts\footnote{Available at https://www.eso.org/sci/facilities/lasilla/\\instruments/feros/tools/DRS.html}. The median peak $S/N$ per 1.4~\kms~spectral pixel is 261.

\subsection{TESS photometry}

TESS is a space telescope obtaining high-precision ($\mu$mag) photometry \citep{2015JATIS...1a4003R}. Its initial mission will last two years, during which it will observe 85\% of the sky in overlapping sectors of $96 \times 24$ deg. Each sector is observed for about 27 days. Data for high-priority targets is downloaded with a 2-minute cadence. The instrument obtains data over a broad bandpass (6000 \AA~to 10,000 \AA), with large ($21 \times 21$ arcsecond) pixels.

HD\,156424 was observed by TESS in Sector 12 using 2-min cadence, with a total of 15,784 individual observations. We obtained the light curve from the Mikulski Archive for Space Telescopes (MAST), selecting the {\sc pdcsap} flux as the light curve with the best apparent detrending. No additional detrending was required, since we are not interested in long-term trends. While the star lies in a relatively crowded field and there are certainly other stars contaminating the TESS light curve, it is the brightest star within about 4 arcminutes. 

\section{Photometric analysis}\label{sec:lc}

\begin{table}
\centering
\caption[]{Frequencies from the TESS light curve and radial velocity measurements, with uncertainties in the last digit given in parantheses. TESS amplitudes are in mmag, with an uncertainty of about 0.01 mmag; RV amplitudes are in \kms, with an uncertainty of about 0.1 \kms. The third column gives the signal-to-noise ratio ($S/N$) of the frequency. The final column gives the identification.}
\label{freqtab}
\begin{tabular}{l r r r l}
\hline\hline
Label & Frequency (${\rm c~d^{-1}}$) & Amplitude & $S/N$ & ID \\
\hline
\multicolumn{5}{c}{TESS} \\
$f_{1}$ &  11.20672(7) & 3.27 & 254 &  \\
$f_{2}$ &  0.71699(8) & 3.10 & 24 & \\
$f_{3}$ &  13.7753(1) & 2.11 & 177 &  \\
$f_{4}$ &  10.8634(1) & 1.75 & 130 & \\
$f_{5}$ &  1.4291(7) & 0.35 & 4 & $ 2f_{2}$ \\
$f_{6}$ &  11.5015(8) & 0.30 & 22 & $2f_1 - f_4$ \\
$f_{7}$ &  22.363(1) & 0.21 & 18 &  $2f_1$ \\
$f_{8}$ &  24.639(3) & 0.06 & 6 &  $f_3 + f_4$ \\
$f_{9}$ &  16.187(4) & 0.06 & 5 &  \\
$f_{10}$ &  24.991(4) & 0.05 & 5 &  $f_1 + f_3$ \\
$f_{11}$ &  22.406(4) & 0.05 & 5 & $2f_1$  \\
\hline
\multicolumn{5}{c}{FEROS} \\
$f_1$ & 11.213(7) & 3.3 & 12 & \\
\hline
\multicolumn{5}{c}{ESPaDOnS} \\
$f_1$ & 11.2067(2) & 4.0 & 28 & \\
\hline
\multicolumn{5}{c}{HARPSpol} \\
$f_1$ & 11.17(1) & 3.4 & 27 & \\
\hline
\multicolumn{5}{c}{Combined spectroscopy} \\
$f_1$ & 11.20691(1) & 3.7 & 33 & \\
$f_2$ & 12.22456(5) & 0.8 & 7 & \\
\hline\hline
\end{tabular}
\end{table}

   \begin{figure}
   \centering
   \includegraphics[width=8.5cm]{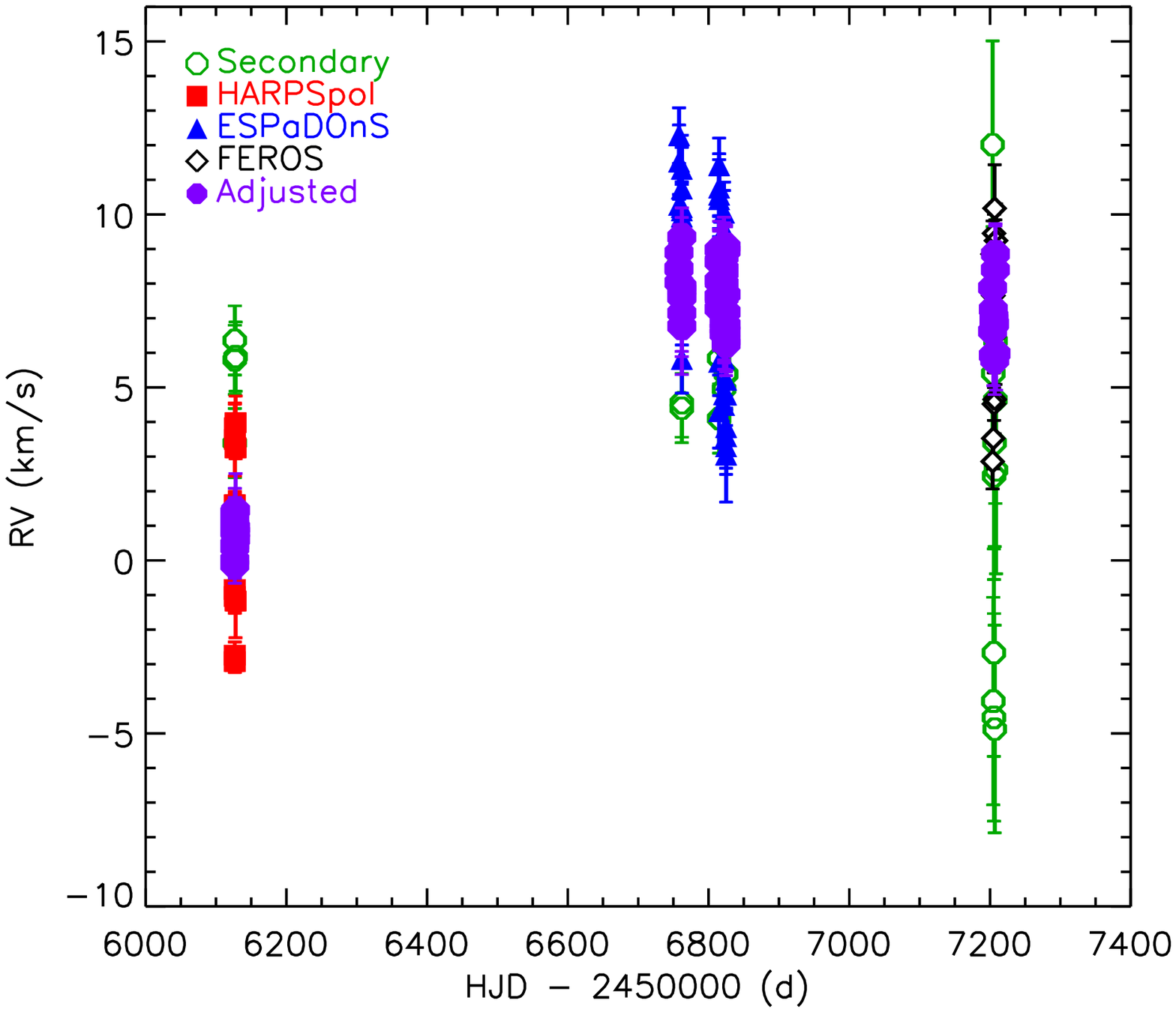} \\
   \includegraphics[width=8.5cm]{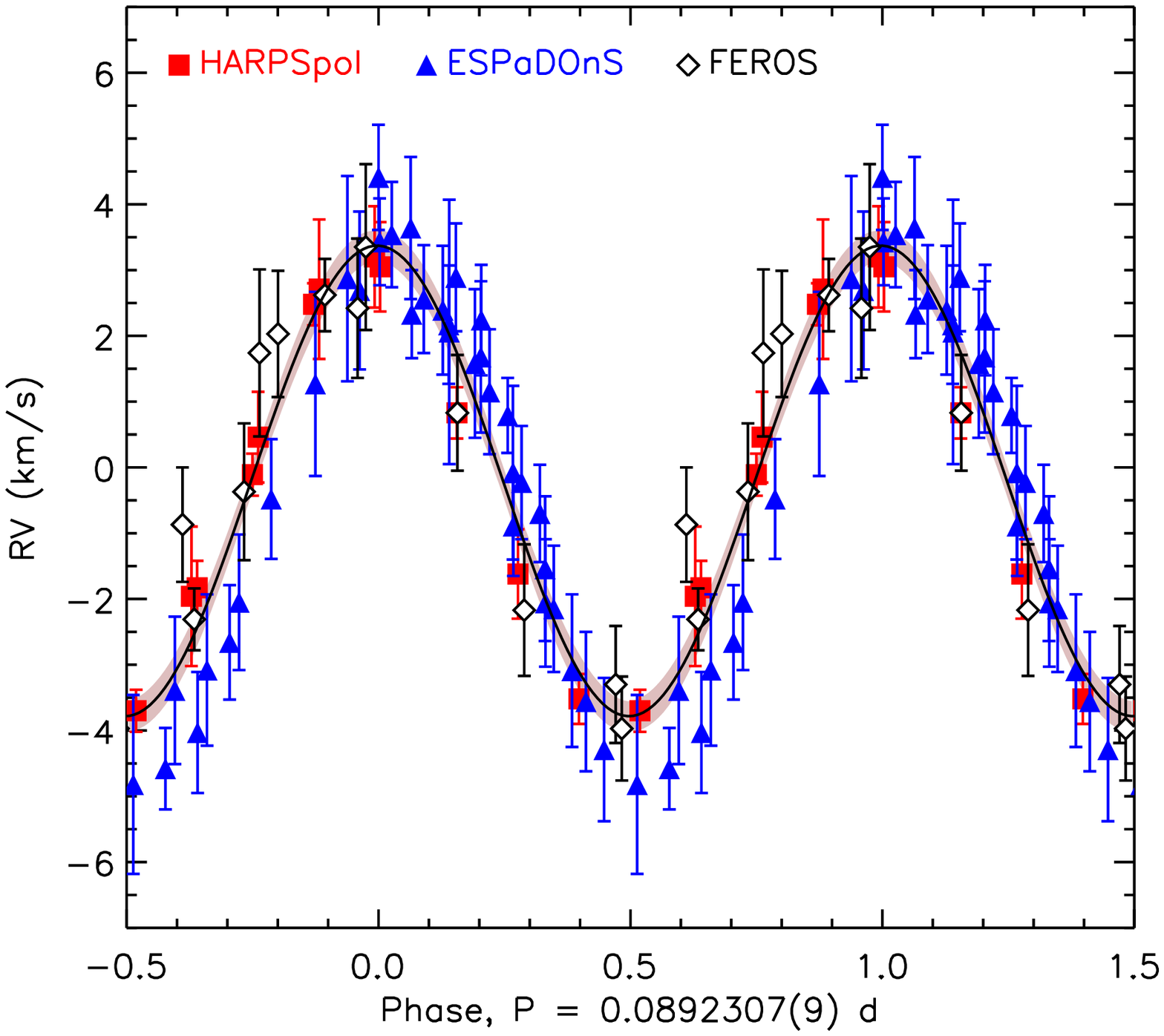} \\
      \caption[]{{\em Top}: RVs obtained from HARPSpol, ESPaDOnS, and FEROS as a function of time. Green points show RV measurements of the secondary obtained from LSD profiles. Purple points show RV measurements of the primary with the pulsational variability (below) removed. While the sharp-lined primary component shows evidence for a long-term RV variation suggestive of binarity, the broad-lined component is consistent with no RV variability. {\em Bottom}: RVs of the primary phased with $f_1$ (top). The curved line and shaded regions show the sinusoidal fit and uncertainties. RVs have been adjusted to the mean value in each dataset in order to remove the long-term variation.}
         \label{hd156424_rv}
   \end{figure}

   \begin{figure}
   \centering
   \includegraphics[width=0.45\textwidth]{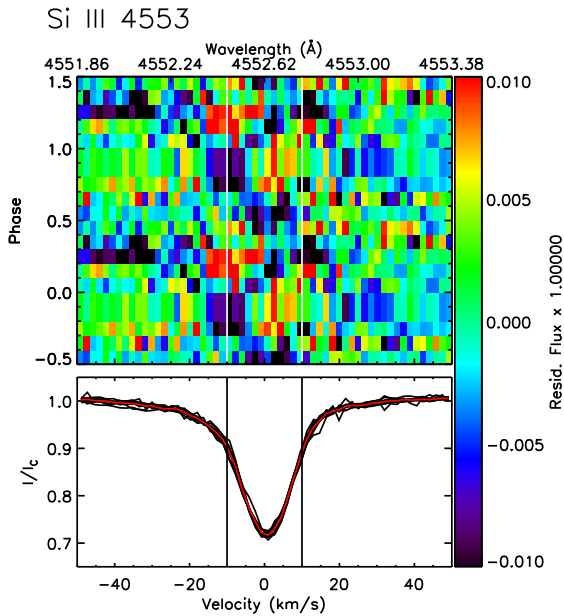}
      \caption[]{Dynamic spectrum of the Si {\sc iii} 4553 line. Individual observations were shifted to the rest velocity, and the mean spectrum was used as the reference spectrum. Solid vertical lines indicate $\pm$\vsini. Residual flux is folded with $f_1$.}
         \label{hd156424_SiIII4553_dyn}
   \end{figure}

   \begin{figure*}
   \centering
   \includegraphics[width=.95\textwidth]{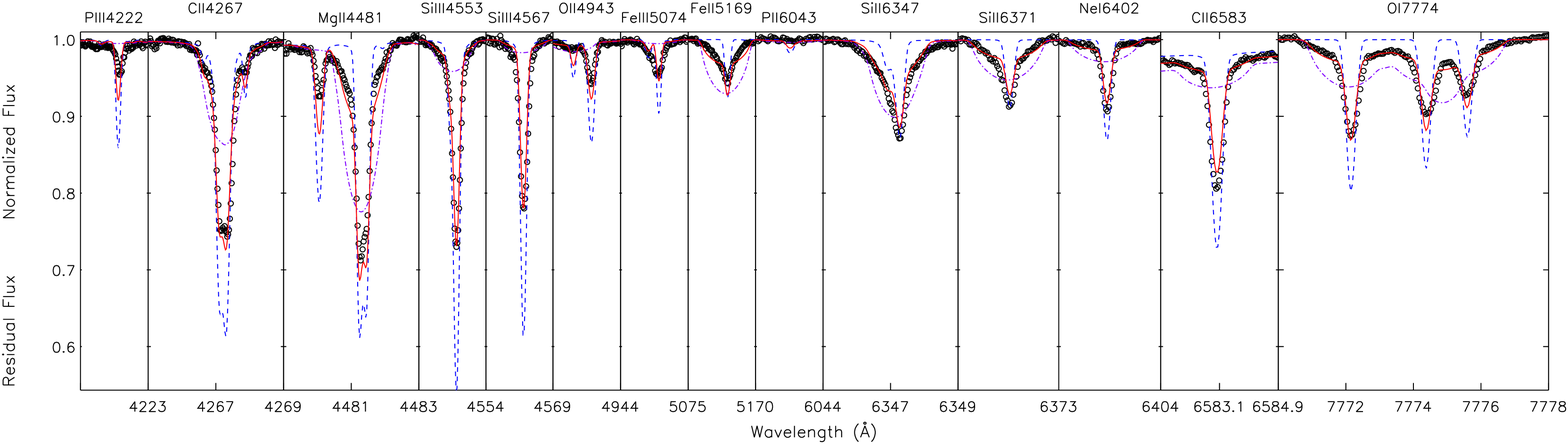} 
   \includegraphics[width=.95\textwidth]{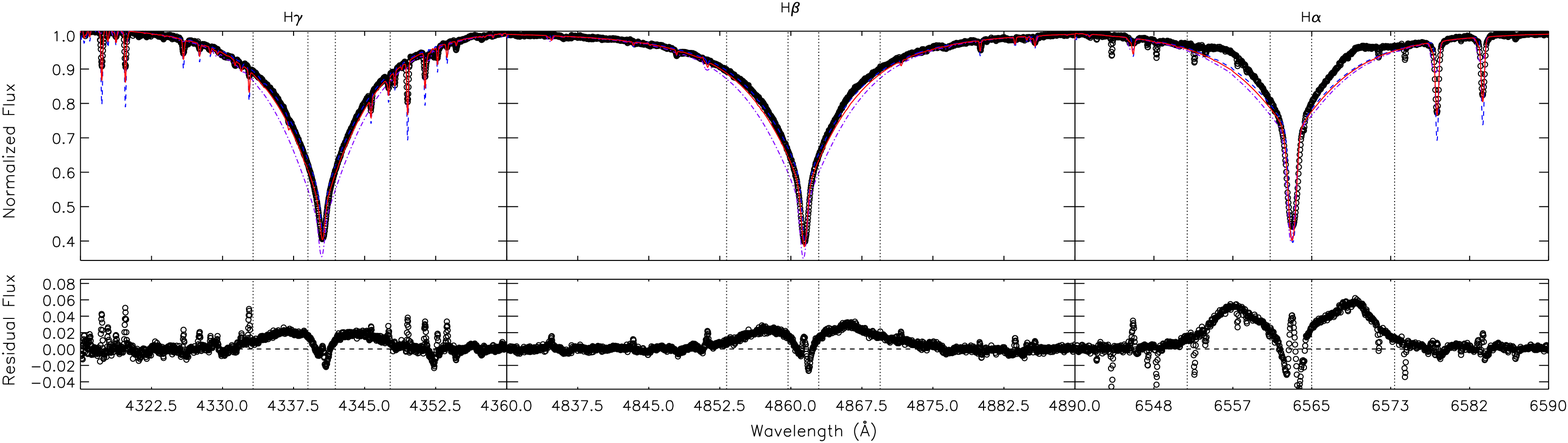} 
      \caption[]{{\em Top}: Synthetic TLUSTY spectrum fits to metallic lines in the mean ESPaDOnS spectrum. Primary in dashed blue, secondary in dot-dashed purple, and combined in solid red. Primary and secondary spectra are intrinsic, i.e.\ they have not been scaled by luminosity. Note the difference in line profile shape between low ionization lines (O~{\sc i}, Si~{\sc ii}, Fe~{\sc ii}) with stronger contributions from the secondary, and high ionization lines (O~{\sc ii}, Si~{\sc iii}, Fe~{\sc iii}) with stronger contributions from the primary. {\em Bottom}: Synthetic TLUSTY spectrum fits to Balmer lines in the mean ESPaDOnS spectrum. Vertical dotted lines delineate regions excluded from the fit due to the presence of circumstellar emission. Bottom panels show residual flux; note the presence of emission in all 3 lines.}
         \label{hd156424_binspec_compare}
   \end{figure*}

The TESS light curve is shown in the right panel of Fig.\ \ref{hd156424_tess_lc} and evidences clear, multi-periodic variability. Analysis of the light curve with {\sc period04} \citep{2005CoAst.146...53L} reveals several significant frequencies. These are listed in Table \ref{freqtab}. The four strongest frequencies have amplitudes of about 2 to 3 mmag, and are indicated in the top left panel of Fig.\ \ref{hd156424_tess_lc}. After pre-whitening with these frequencies, 7 more frequencies with amplitudes of about 0.1 mmag are detected; these are shown in the bottom left panel of Fig.\ \ref{freqtab}. Significance was determined according to the usual criterion of a $S/N$ of at least 4 \citep{1993A&A...271..482B,1997A&A...328..544K}. This noise floor is shown in Fig.\ \ref{hd156424_tess_lc}, where it was determined by fitting a low-order polynomial to the pre-whitened frequency spectrum in log-log space. 

The majority of the frequencies are above 10 ${\rm d}^{-1}$, and are consistent with $p-$mode pulsations. Three of the $p-$modes ($f_1$, $f_4$, and $f_6$), including the strongest frequency, constitute a triplet centered on $f_1$, with a separation of about 0.3~${\rm d}^{-1}$. With the exception of $f_9$, the remaining high-frequency terms all appear to be either harmonics or linear combinations of the strongest frequencies, as determined using the Rayleigh criterion with a full-width-half-max of about 0.06 c/d.

In addition to the $p-$modes, there are two low-frequency terms in the frequency spectrum, the lowest of which, $f_2$, has the second-highest amplitude of all detected frequencies. The weaker signal at $f_4$ is very close to the first harmonic of the stronger low-frequency term $f_2$, suggesting that this might be due to rotational modulation. The period corresponding to $f_2$, about 1.4~d, is about half the period reported by \cite{2018MNRAS.475.5144S}. This could indicate that $f_2$ is in fact the first harmonic of $f_{\rm rot}$; however, there is no statistically significant peak at the corresponding frequency. Therefore, if $f_2 = 2f_{\rm rot}$, the rotationally modulated light curve would need to be an almost perfect double-wave variation. It is interesting to note that there are two sets of frequencies ($f_1 - f_4$ and $f_8 - f_{10}$) that are separated by about 0.35 ${\rm d}^{-1}$, which is very close to the presumptive $f_{\rm rot} = f_2 / 2$. The elements of the triplet centred on $f_1$ are also separated by about 0.3~${\rm d}^{-1}$. These frequency groups may therefore be a consequence of frequency splitting, supporting the interpretation of rotational modulation for $f_2$. On the other hand, this could be coincidental, and the low-frequency terms might be due to $g$-mode pulsations.

\section{Radial velocities}\label{sec:rv}

Radial velocities (RVs) were measured from individual unpolarized spectra, yielding 36 RV measurements from the ESPaDOnS data, 12 from the FEROS data, and 12 from the HARPSpol data. RVs were measured via the centre-of-gravity of the Si {\sc iii} $\lambda\lambda$ 4553, 4568, and 4575 lines, as well as the O {\sc ii} $\lambda\lambda$ 4415 line. As explained below in \S~\ref{sec:multiplicity}, these lines are dominated by the spectrum of the primary. All yielded similar results within uncertainties. The weighted mean across all 4 lines was then taken so as to increase the $S/N$. RVs are tabulated in Table \ref{rvtab}.

RVs are shown as a function of time in the top panel of Fig.\ \ref{hd156424_rv}. It is immediately apparent that there is a systematic difference between the HARPSpol RVs and the RVs measured from the ESPaDonS and FEROS datasets, with a difference of about 7~\kms. The most natural explanation for this is orbital motion due to the influence of a companion star. 

Fourier analysis of the ESPaDOnS RVs using {\sc period04} yielded a single significant frequency, $11.2067(2)~{\rm d}^{-1}$, with a $S/N$ of 28. After pre-whitening with this frequency, the next highest peak has a $S/N$ below the significance threshold of 4. The FEROS measurements yield $11.213(7)~{\rm d}^{-1}$, with a $S/N$ of 12. The HARPSpol data yields 11.17(1)~${\rm d}^{-1}$, with a $S/N$ of 27. Combining the three datasets (after removing the mean RV of each dataset so as to correct for the systematic differences) yields $11.20691(1)~{\rm d}^{-1}$. A second frequency is found at 12.22456(5) ${\rm d}^{-1}$, although the absence of this frequency in the much more precise photometric dataset suggests it may be spurious (it is also worth noting that the amplitude is similar to the RV uncertainty). RVs are shown phased with $f_1$ (as determined from RVs) in the bottom panel of Fig.\ \ref{hd156424_rv}. 

In $\beta$ Cep pulsators, photometric variations occur primarily due to changes in \teff~with pulsation phase, which can sometimes also be detected as changes in line strength. Fig.\ \ref{hd156424_SiIII4553_dyn} shows a dynamic spectrum for Si~{\sc iii}~$\lambda\lambda$~4553, with individual spectra moved to the laboratory frame and folded with $f_1$. With RV variation removed, there is no apparent line profile variability down to about 1\% of the continuum. This is probably consistent with the very low photometric and RV amplitude of $f_1$, since intrinsic variation in line profile strength is due to the change in surface temperature, which is minimal in this case.

\section{Multiplicity}\label{sec:multiplicity}


   \begin{figure}
   \centering
   \includegraphics[width=.45\textwidth]{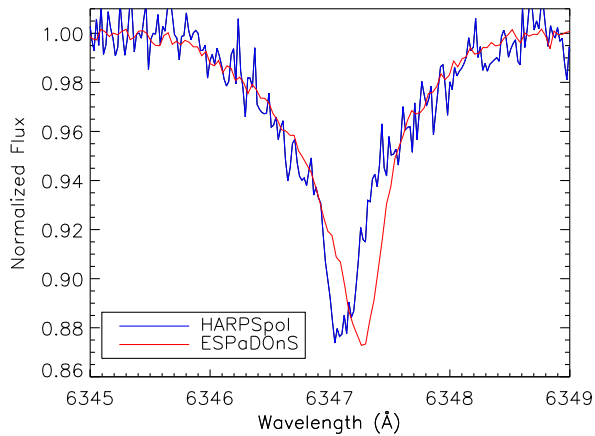} 
      \caption[]{The Si~{\sc ii} 6347 line in the mean HARPSpol and ESPaDOnS spectra. In each line profile a narrow-lined component (corresponding to HD\,156424A) and a broad-lined component (HD\,156424B) are distinguishable. The narrow-lined component is redshifted in the ESPaDOnS observation as compared to the HARPSpol observation; the RV of the broad-lined component is, however, stable.}
         \label{hd156424_SiII6347}
   \end{figure}

A companion star is known to be about $0.35''$ away \citep{1993AJ....106..352H,2010AJ....139..743T}. Sco OB4 is estimated to be at a distance of $1300^{+500}_{-200}$ pc \citep{2005AA...438.1163K}, so the projected separation of the stars is 455$^{+175}_{-70}$~AU. The magnitude difference is estimated at 2.3 mag in the $y$ band \citep{2010AJ....139..743T}, or about a factor of 8 in luminosity (implying the companion star should have $\log{L} \sim 2.8$). In this case it should have a mass of about 5 or 6 \msun, for a mass ratio of $\sim$3. From Kepler's third law the orbital period would then need to be on the order of 2000 years. However, speckle observations indicate the companion's position angle changed by 30$^\circ$ between 1990 and 2008 \citep{1993AJ....106..352H,2010AJ....139..743T}; this would suggest an orbital period closer to 200 years. In either case this is much too long to be consistent with the observed long term RV variation.





\cite{alecian2014} indicated that they found no indication of the companion star in the spectrum. However, given the companion star's angular distance from the primary, it was inside the $2''$ FEROS aperture, the $1''$ HARPSpol aperture, and the $1.8''$ ESPaDOnS pinhole. To conduct a more detailed investigation, we created a mean spectrum from all available ESPaDOnS spectra, achieving a peak $S/N$ per 1.8 \kms~pixel of about 800. We then examined a selection of metallic lines, including especially lines for which different ionizations are available in the spectrum. These spectral lines are shown in the top panels of Fig.\ \ref{hd156424_binspec_compare}. Since the companion star is estimated to be about 2.3 mag dimmer than the primary, it should have a significantly lower \teff, and should therefore contribute different amounts to the flux of lines with different ionizations. Comparing O~{\sc i} to O~{\sc ii}, Fe~{\sc ii} to Fe~{\sc iii}, and Si~{\sc ii} to Si~{\sc iii}, this pattern is clearly apparent. Lines with lower ionizations have much more extended wings than lines of higher ionizations. 

The top panels of Fig.\ \ref{hd156424_binspec_compare} show a fit to a selection of metallic lines in the mean ESPaDOnS spectrum using synthetic {\sc tlusty} spectra from the BSTAR2006 library \citep{lanzhubeny2007}. The fitting was performed using a grid-based search, covering $20 {\rm kK} < T_{\rm eff,A} < 24~{\rm kK}$ for the primary, $15 {\rm kK} < T_{\rm eff,B} < 20~{\rm kK}$, $3.5 < \log{g} < 4.5$ for both stars, $1 < v\sin{i}_{\rm A} < 10$~\kms, and  $10 < v\sin{i}_{\rm B} < 100$~\kms. The radius ratio $R_{\rm A}/R_{\rm B}$ was allowed to vary as a free parameter, and was calculated using the continuum fluxes from the TLUSTY spectra. RVs were fixed to 8 and 5 \kms~for A and B, respectively. The best fit was obtained for $\log{g} = 4.25$ for both stars, $T_{\rm eff,A} = 23 \pm 1$~kK, $T_{\rm eff,B} = 16 \pm 1$~kK, $v\sin{i}_{\rm A} = 4.4 \pm 1.5$~\kms, and $v\sin{i}_{\rm B} = 25 \pm 2$~\kms. The best-fit value of $R_{\rm A}/R_{\rm B} = 1.1$ certainly over-estimates the radius of secondary, which may be a consequence of chemical peculiarities in one or both stars. This results in HD\,156424A being about twice as bright as HD\,156424B at visible wavelengths, whereas the magnitude difference estimated by \cite{2010AJ....139..743T} implies the primary should be about 8 times brighter than the secondary; however, \citeauthor{2010AJ....139..743T} noted that the magnitude difference is probably over-estimated due to the noisy data.

The bottom panels of Fig.\ \ref{hd156424_binspec_compare} show fits to the H Balmer lines H$\gamma$, H$\beta$, and H$\alpha$. In this case the $T_{\rm eff}$ of the two components were fixed to their best-fit values. Due to the presence of obvious circumstellar emission in H$\alpha$, the regions with emission were excluded from the fit; the corresponding regions in H$\gamma$ and H$\beta$ were also excluded. The best fit is obtained for $\log{g}_{\rm A} = 4.0$ and $\log{g}_{\rm B} = 3.75$. This is a curious result since, assuming the two stars to be coeval, the surface gravity of HD\,156424B should be at least as high, or higher, than that of the more massive primary. However, if the grid is restricted to force $\log{g}_{\rm A} < \log{g}_{\rm B}$, the difference in fit quality is negligible, and it seems likely that, given the small radial velocity separation of the components, their respective surface gravities cannot be confidently disentangled.



RV measurements of the secondary were conducted using the parameterized line profile fitting program described by \cite{2017MNRAS.465.2432G}. These are shown in Fig.\ \ref{hd156424_rv}, and indicate that HD\,156424B's RV is stable between the HARPSpol and ESPaDOnS datasets. There is apparent variation within the FEROS dataset, however this is almost certainly scatter caused by the secondary's intrinsic line profile variability, which is more apparent in the FEROS dataset. The absence of RV variation in HD\,156424B is clearly apparent from a comparison of the Si~{\sc ii} $\lambda\lambda$ 6347 line between the mean HARPSpol and ESPaDOnS spectra (Fig.\ \ref{hd156424_SiII6347}). Since HD\,156424B is the less luminous and therefore presumably less massive of the two stars, if it were responsible for HD\,156424A's long term RV variation it would of necessity have a larger RV amplitude. It thus seems likely that the primary's long term RV variation is caused by an undetected third star, and that the system is a hierarchical triple, with HD\,156424A and the undetected companion in an orbit with a period of $\sim$years, and HD\,156424B orbiting the inner pair with an orbital period of centuries (as inferred from visual data). In this scenario, HD\,156424B is the distant companion previously identified by \cite{1993AJ....106..352H} via speckle interferometry. 

A lower limit can be placed on the mass of the undetected companion if we assume that the observed RV variation samples about half of an orbital period (in which case the orbital period would be about 5 years), and the semi-amplitude of the orbital RV variation is therefore about 4 \kms. Half a period is assumed because the mean RV had not yet returned to the HARPSpol value at the time of the FEROS observations; at the same time, the lack of variation between the ESPaDOnS and FEROS data indicates a low rate of change through this time span, suggesting the RV curve is near the maximum of an approximately sinusoidal variation. We further assume that HD\,156424A has a mass of about 8 \msun, as would be appropriate for a young star with its \teff. By iteratively solving for the mass of the unseen companion using the mass function of the system (e.g., \citealt{1973bmss.book.....B}; see also Eqns. A-1 and A-2 in \citealt{2020A&A...637L...3R}), if the orbital inclination is 90$^\circ$ and the eccentricity is 0 the companion must have a mass of at least 1 \msun. An eccentric orbit would indicate a lower mass for the companion. Since a 1 \msun~star would have a bolometric luminosity of less than a thousandth that of HD\,156424A, it is entirely possible for such a star to remain undetected. An approximate upper limit on the mass of the companion can be placed under the assumption that it would be seen if its mass were comparable to that inferred for HD\,156424B from its \teff~at the ZAMS, about 6 \msun; in this case, the orbital inclination must be greater than 13$^\circ$.

There is an apparent change in $f_1$ between the combined RVs and the TESS light curve. This is significant at almost the 3$\sigma$ level using the larger TESS uncertainty. The change amounts to $\Delta P \sim 1.5 \times 10^{-6}$~d. If this is due to the light-time effect, there should be a corresponding change in RV of $\Delta{\rm RV} = c\Delta P/P \sim 5$~\kms~\citep{1992A&A...261..203P}. Referring to Fig.\ \ref{hd156424_rv}, this is about the change in RV that is observed between the acquisition of the HARPSpol and ESPaDOnS/FEROS datasets. The apparently much larger difference in $f_1$ as determined from the HARPSpol dataset and others is also formally significant at the 3$\sigma$ level using the larger HARPSpol uncertainty, but would require $\Delta RV \sim 1000$~\kms~to be explained via the light-time effect. This change, if real, therefore cannot be explained by the light-time effect; however, as it is based on a relatively small number of measurements that are phased equally well using a period closer to the mean value, this large difference in frequencies is probably spurious.  

\section{Magnetometry}\label{sec:mag}

   \begin{figure}
   \centering
   \includegraphics[width=8.5cm]{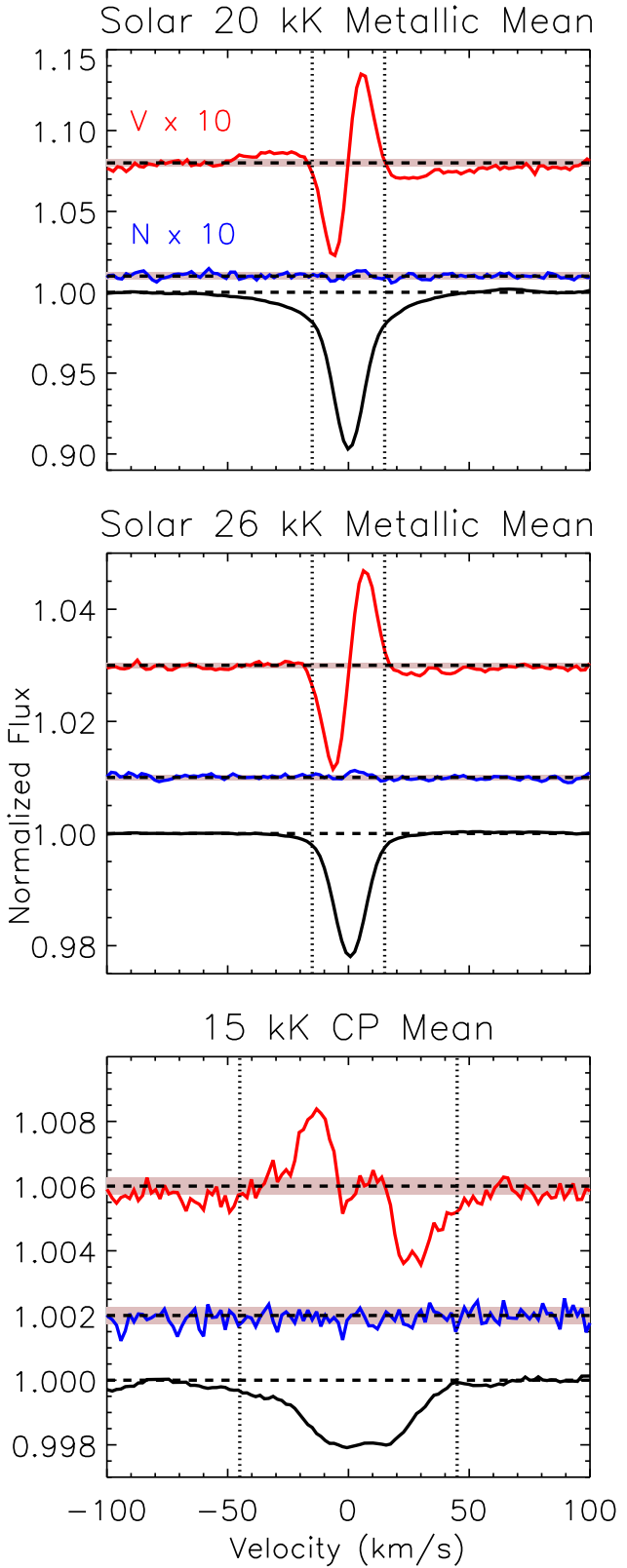} 
      \caption[]{Mean LSD profiles extracted with various line masks. The grey shaded regions indicate the mean uncertainties in $N$ and Stokes $V$.}
         \label{hd156424_lsd_all}
   \end{figure}

   \begin{figure}
   \centering
   \includegraphics[width=8.5cm]{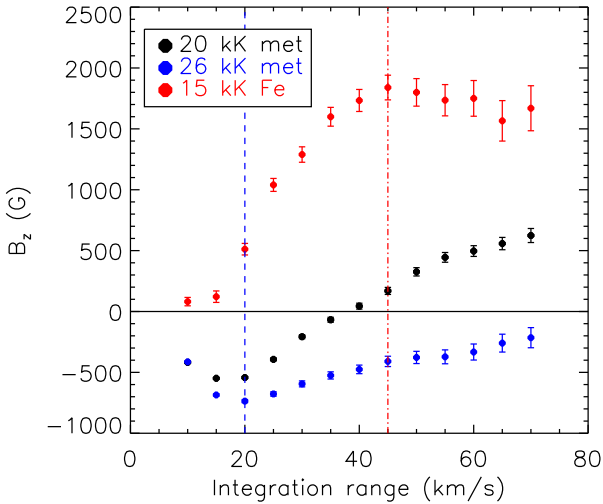} 
      \caption[]{Change in \bz~determined from mean LSD profiles as a function of the integration range width. Vertical dashed blue and dot-dashed red lines indicate the line widths of the A and B components.}
         \label{hd156424_int_bz}
   \end{figure}

   \begin{figure*}
   \centering
\begin{tabular}{cc}
   \includegraphics[width=8.5cm]{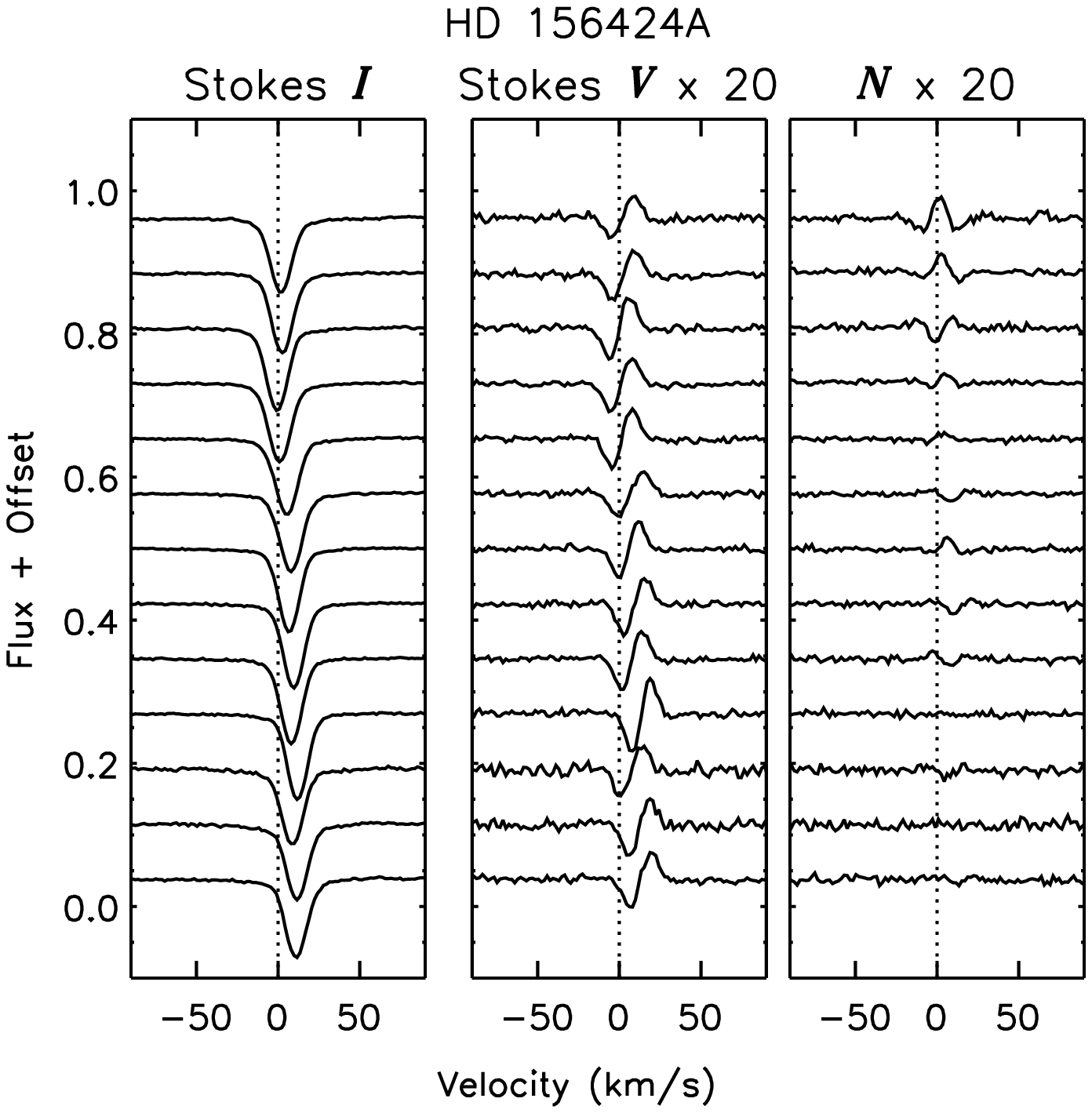} &
   \includegraphics[width=8.5cm]{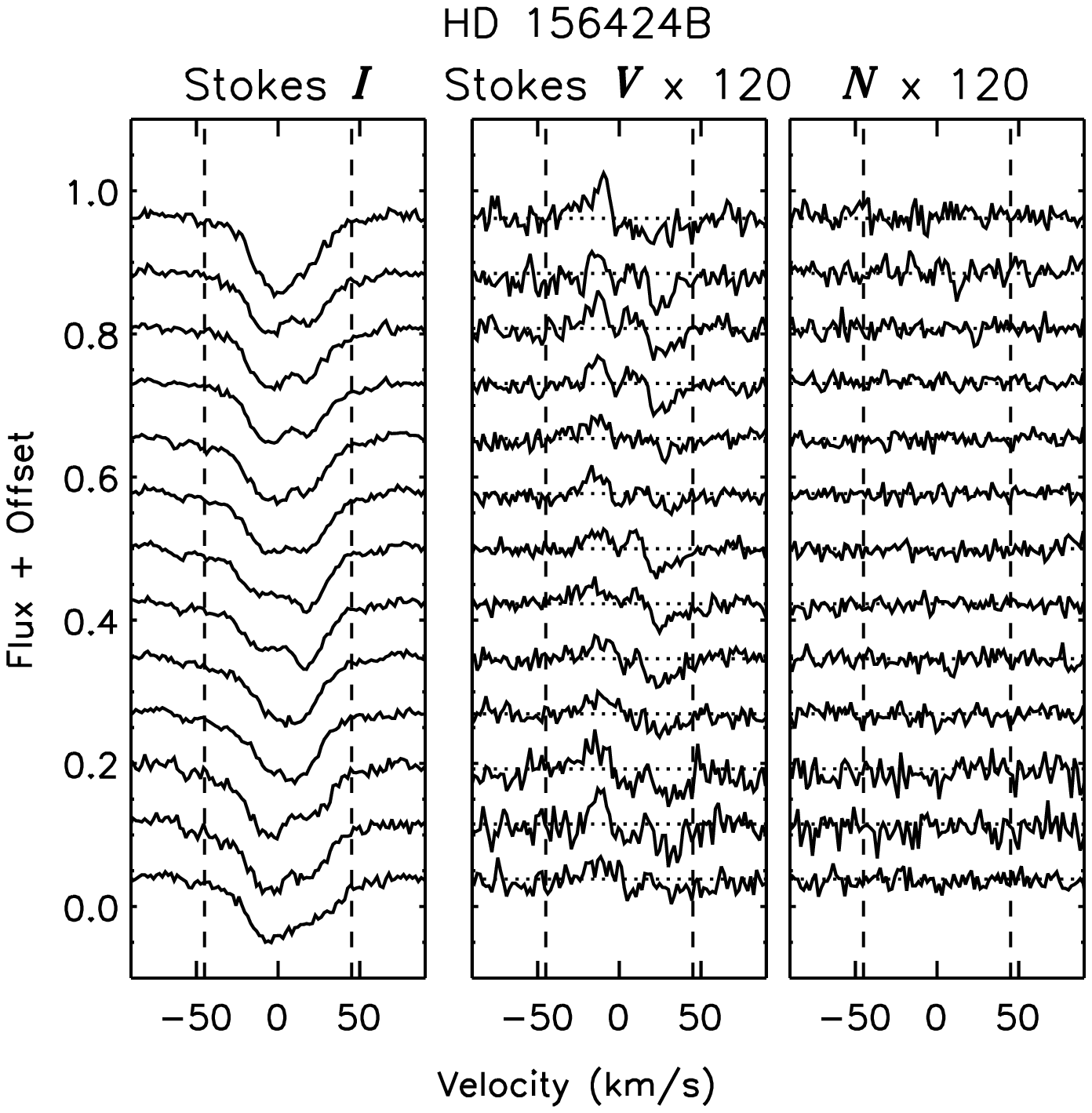} \\
\end{tabular}
      \caption[]{Individual LSD profiles for HD\,156424A (left) and HD\,156424B (right). Profiles are organized in temporal order, with the first observation on the bottom. The vertical dashed line in the left panels indicates the mean radial velocity for HD\,156424A. Note that the Stokes $V$ profiles track the RV variation of Stokes $I$. Note also the clear signatures in several of the $N$ profiles. Vertical dashed lines in the left panels indicate the limits of the Stokes $I$ profile of HD\,156424B. Note the lack of RV variation in either Stokes $I$ or $V$. Asymmetry in Stokes $I$ does not appear to be related in an obvious way to RV variation of HD\,156424A's Stokes $I$ profiles, suggesting that the line profile variability of HD\,156424B is not a consequence of contamination by the Stokes $I$ profile of HD\,156424A.}
         \label{hd156424_lsd_phaseplot}
   \end{figure*}

\begin{table*}
\centering
\caption[]{\bz~and \nz~measurements for the two stellar components. The first column indicates the instrument with which the measurement was obtained: H(ARPSpol) or E(SPaDOnS). `DF' refers to `detection flag': either definite detection (DD), marginal detection (MD), or non-detection (ND).}
\label{bztab}
\begin{tabular}{l r r r l r l r l r l}
\hline\hline
 & & & \multicolumn{4}{c}{Primary} & \multicolumn{4}{c}{Secondary} \\
Inst. & HJD -- & Date & \bz & DF$_V$ & \nz & DF$_N$  & \bz & DF$_V$ & \nz & DF$_N$ \\
  & 2456000 &   & (G) &  & (G) &   & (G) &  & (G) &  \\
\hline
H & 126.66421 & 18/07/2012 & $-849\pm26$ &  DD & $-7\pm26$ &  DD & $2154\pm274$ &  DD & $415\pm270$ &  ND \\
H & 126.70743 & 18/07/2012 & $-901\pm38$ &  DD & $-48\pm37$ &  DD & $2586\pm391$ &  DD & $-114\pm385$ &  ND \\
H & 127.77721 & 19/07/2012 & $-686\pm48$ &  DD & $-4\pm48$ &  DD & $3005\pm392$ &  DD & $-403\pm385$ &  ND \\
E & 758.03881 & 10/04/2014 & $-846\pm54$ &  DD & $4\pm53$ &  ND & $1645\pm418$ &  ND & $249\pm416$ &  ND \\
E & 761.95376 & 14/04/2014 & $-801\pm79$ &  DD & $43\pm79$ &  ND & $1952\pm528$ &  ND & $-415\pm525$ &  ND \\
E & 761.97745 & 14/04/2014 & $-742\pm74$ &  DD & $0\pm74$ &  ND & $3053\pm541$ &  ND & $-978\pm534$ &  ND \\
E & 814.96460 & 06/06/2014 & $-867\pm43$ &  DD & $10\pm43$ &  ND & $1963\pm373$ &  MD & $-43\pm370$ &  ND \\
E & 814.98761 & 06/06/2014 & $-821\pm37$ &  DD & $45\pm37$ &  DD & $2281\pm337$ &  DD & $-326\pm333$ &  ND \\
E & 821.94039 & 13/06/2014 & $-818\pm36$ &  DD & $-8\pm36$ &  DD & $2777\pm335$ &  DD & $-79\pm328$ &  ND \\
E & 821.98111 & 13/06/2014 & $-769\pm35$ &  DD & $26\pm35$ &  DD & $1801\pm315$ &  DD & $-2\pm312$ &  ND \\
E & 824.89544 & 16/06/2014 & $-859\pm33$ &  DD & $7\pm33$ &  DD & $1234\pm266$ &  DD & $-128\pm265$ &  ND \\
E & 824.91836 & 16/06/2014 & $-853\pm34$ &  DD & $-3\pm34$ & DD & $1533\pm393$ & DD & $-280\pm379$ & ND \\
\hline\hline
\end{tabular}
\end{table*}



In order to measure the magnetic field with the maximum possible precision, mean line profiles were extracted from the spectropolarimetric data using Least Squares Deconvolution \citep[LSD;][]{d1997}, for which we employed the iLSD package \citep{koch2010}. Line masks were obtained from the Vienna Atomic Line Database \citep[VALD3;][]{piskunov1995,ryabchikova1997,kupka1999,kupka2000,2015PhyS...90e4005R} using `extract stellar' requests. The first such line mask was obtained for the stellar parameters of HD\,156424 (\teff~$=20$~kK, $\log{g} = 4$) reported by \cite{2019MNRAS.485.1508S}, with a normalized line depth threshold of 0.1. The line mask was then prepared in the usual way: cleaned of contaminating H lines, He lines with broad wings, telluric lines, and interstellar lines (including 2 Diffuse Interstellar Bands at 5780 \AA and 6604 \AA), and the line strengths adjusted to match the observed line depths \citep[e.g.][]{2018MNRAS.475.5144S}. 

The presence of a magnetic field was evaluated using False Alarm Probabilities \citep{dsr1992,d1997}. All Stokes $V$ profiles yield `Definite Detections', i.e.\ FAP~$<10^{-5}$. However, 8/12 $N$ profiles also yield DDs, the remainder being non-detections (NDs; FAP~$>10^{-3}$). These detections in $N$ are due to the high-frequency RV variation from $\beta$ Cep pulsations. Subexposure times for the ESPaDOnS data are about 6\% of $f_1$, and total spectropolarimetric sequence times are 23\% of $f_1$. HARPSpol sequences, all 3 of which show signatures in $N$, span 47\% of a pulsation cycle. This phenomenon has been reported in other stars with rapid RV variation, e.g.\ the $\beta$ Cep stars HD\,96446 and $\xi^1$ CMa \citep{neiner2012a,2017MNRAS.471.2286S}, and the short-period binary HD\,156324 \citep{2018MNRAS.475..839S}. While Stokes $V$ is undoubtedly also affected, the total strength of the line-of-sight disk-integrated magnetic field \bz~\citep[defined by][]{mat1989} should not be affected, as determined via correction of supexposures for RV variation \citep{neiner2012a} and via direct modelling of Stokes $I$, Stokes $V$, and $N$ \citep{2017MNRAS.471.2286S}.

Close examination of the Stokes $V$ profiles, however, reveals a curious feature: lobes of net polarization extending outside of the line profile (Fig.\ \ref{hd156424_lsd_all}, top). To investigate the effect of these anomalous polarization lobes, \bz was measured from the mean line profile using progressively wider integration ranges. In general, \bz~should change up to the edge of the line width, following which it should stabilize while the uncertainty continues to grow \citep[e.g.][]{neiner2012b}. As can be seen in Fig.\ \ref{hd156424_int_bz}, for the LSD profiles extracted using the 20 kK mask \bz~continues to change outside of the error bars well after this point, and indeed reverses sign around 40~\kms. This behaviour is quite anomalous and cannot be explained as a consequence of pulsational influence on the line profile, which should only be able to shift the Stokes $V$ profile by a few \kms. 

The lobes in Stokes $V$ correspond to the very broad wings of Stokes $I$. Since He lines were excluded from the line mask, the presence of these wings cannot be explained as due to Stark broadening. The `wings' may instead be explained by the contribution of of HD\,156424B to the spectrum.


The strange behaviour of \bz, and the correspondance between the `wings' in Stokes $I$ and the lobes in Stokes $V$, suggest that HD\,156424B may possess its own magnetic field and therefore contaminates both Stokes $I$ and $V$. Given its likely \teff~of about 16 kK, if the star is indeed magnetic it is almost certainly a He-weak Bp star. A line list was downloaded from VALD using enhanced abundances (${\rm Si=-4}$, ${\rm Cl=-5}$, ${\rm Ti=-5.5}$, ${\rm Cr=-5}$, ${\rm Fe=-3.8}$, ${\rm Ni=-5.3}$, ${\rm Ba=-7.8}$, ${\rm Ce=-6}$, ${\rm Pr=-8.2}$, and ${\rm Nd=-7.4}$, chosen using the mean surface abundances for HR 2949 \citep{2015MNRAS.449.3945S}, which has a similar \teff). To emphasize lines formed at lower \teff~and therefore hopefully help to separate the signals, a \teff~of 15 kK was used. The mask was then cleaned in the usual fashion to remove contamination from H Balmer, He, interstellar, and telluric lines. The mask was further cleaned so as to include only low-ionization (mostly Fe~{\sc ii}) lines, i.e.\ spectral lines in which the contribution from HD\,156424B is dominant (see Fig.\ \ref{hd156424_binspec_compare}). Finally, a complementary mask was obtained using a solar abundance 26 kK line mask template, which was cleaned to remove any lines appearing in the 15 kK mask; in this case a higher \teff~than that inferred from modelling of HD\,156424A was chosen so as to emphasize lines appearing at higher ionizations that can be expected to minimize the contribution of the cooler star. LSD profiles were then extracted simultaneously using these two masks, a unique capability of the iLSD package \citep{koch2010}. \bz~measurements for the two sets of LSD profiles are listed in Table \ref{bztab}. \bz~measurements are not affected by continuum dilution in binary star spectra, since they are are normalized using the EW of the Stokes $I$ profile \citep{mat1989}; so long as Stokes $I$ is not contaminated by the other star, the measurement should therefore give an accurate indication of the line-of-sight magnetic field strength of the star in question.

The LSD profiles extracted using the 26 kK mask show a greatly reduced broadening in the wings of Stokes $I$ (Fig.\ \ref{hd156424_lsd_all}, middle), indicating that the contribution of the secondary has been largely removed. There are furthermore much weaker lobes in Stokes $V$, as verified via the integration range test in Fig.\ \ref{hd156424_int_bz} where it can be seen that, while \bz~indeed continues to change outside of HD\,156424A's line profile, it does not change sign.

The LSD profiles extracted from the 15 kK mask, by contrast, uniformly yield positive \bz~of around $1.5$~kG. As can be seen from the bottom panel of Fig.\ \ref{hd156424_lsd_all}, the Stokes $V$ profile obtained with this mask extends across the broader Stokes $I$ profile of HD\,156324B; has an opposite sign; and corresponds neatly to the Stokes $V$ lobes seen in the LSD profile from the original 20 kK mask. The integration test shows that \bz~stabilizes around 50~\kms, the line width of HD 156424 B's Stokes $I$ profile. 8/12 observations furthermore yield definite detections. All indications are therefore that HD\,156424B also possesses a strong magnetic field. 

The variability of the two components' LSD profiles is illustrated in Fig.\ \ref{hd156424_lsd_phaseplot}. Aside from RV variations, the Stokes $I$ profile of HD 156424A is not strongly variable. By contrast, the Stokes $I$ profiles of HD 156424B are clearly variable. While such variation is expected for CP stars, and is furthermore consistent with the evidence for line profile variability in HD\,156424B (see Fig. \ref{vsini}), given that there is almost certainly some degree of blending with HD 156424A's line profiles, it is natural to wonder if the variability is simply caused by the changing RV of HD 156424A. The vertical dotted line in the left-hand panels of Fig.\ \ref{hd156424_lsd_phaseplot} provides a reference point by which to judge the RV of the primary. Comparison of LSD profiles of HD 156424B possessing similar Stokes $I$ profiles, to the LSD profiles for HD 156424A obtained from the corresponding observations, suggests that RV variation is not the cause of the secondary's variability. From bottom to top, the first and last observations have similar Stokes $I$ profiles for HD 156424B but clearly different RVs for HD 156424A. The same is true of the second and third, as compared to the eleventh and twelfth, observations. The line profile variability of HD\,156424B is most likely due to chemical spots. Zeeman splitting is also possible, given the strong \bz~measurements; however this is difficult to verify in LSD profiles.

\bz~measurements show very little variation for either component. For HD 156424A, the highest peak in the \bz~periodogram is at 1.539(2)~d, corresponding to about 0.65~d$^{-1}$; there is no peak in the TESS frequency spectrum corresponding to this frequency. However, the $S/N$ of this peak is below 4, therefore it is not statistically significant. We are furthermore unable to confirm the 2.8 d period determined by \cite{2018MNRAS.475.5144S}, and believe that this period was a spurious consequence of the unidentified contribution of the secondary. If we take the lowest frequency obtained from the TESS light curve, corresponding to a period of about 1.4 d, the \bz~measurements are not satisfactorily phased. We conclude that either a) the lowest-frequency term in the TESS dataset is not, in fact, the rotation period, or b) HD 156424B is still affecting HD 156424A's Stokes $V$ profile and the \bz~measurements are therefore not reliable. The latter conclusion seems more likely in light of the results of the integration range test (Fig.\ \ref{hd156424_int_bz}).

For HD 156424B, the strongest peak in the periodogram is at 0.8773(5) d, or about 1.14~d$^{-1}$. Once again this does not correspond to any of the frequencies in the TESS light curve; however, the $S/N$ of this frequency is below 4, therefore it is probably spurious. Similarly to the case of the primary, the lowest frequencies in the TESS light curve do not satisfactorily phase \bz, and our conclusions for this star are identical. We conclude that $P_{\rm rot}$ cannot be determined using \bz~measurements alone for either star. 


\section{H$\alpha$ emission and rotation}\label{sec:halpha}

   \begin{figure}
   \centering
   \includegraphics[width=8.5cm]{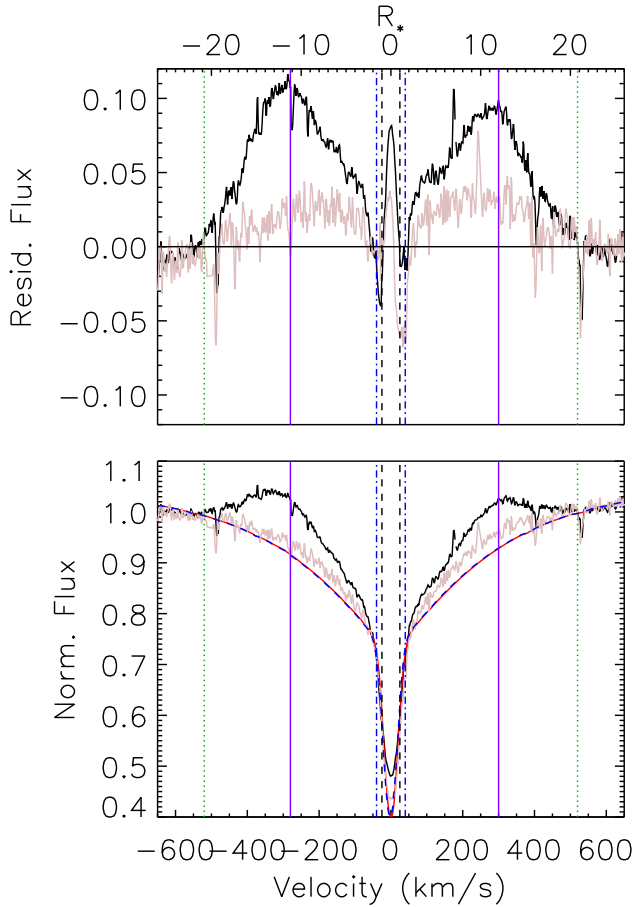} 
      \caption[]{{\em Bottom}: H$\alpha$ profiles at maximum (black) and minimum (gray) emission. Solid red and dashed blue show the respective synthetic binary spectra. Vertical lines: dashed black: $\pm$\vsini~for HD\,156424B; dot-dashed blue: Kepler corotation radius \rk~for HD\,156424B; solid purple: maximum emission; dotted green: zero emission. {\em Top}: Residual flux after subtraction of synthetic H$\alpha$ profiles. The top horizontal scale uses \vsini~for HD\,156424B to obtain projected radius from velocity.}
         \label{hd156424_halpha}
   \end{figure}

   \begin{figure}
   \centering
\begin{tabular}{cc}
   \includegraphics[width=.225\textwidth]{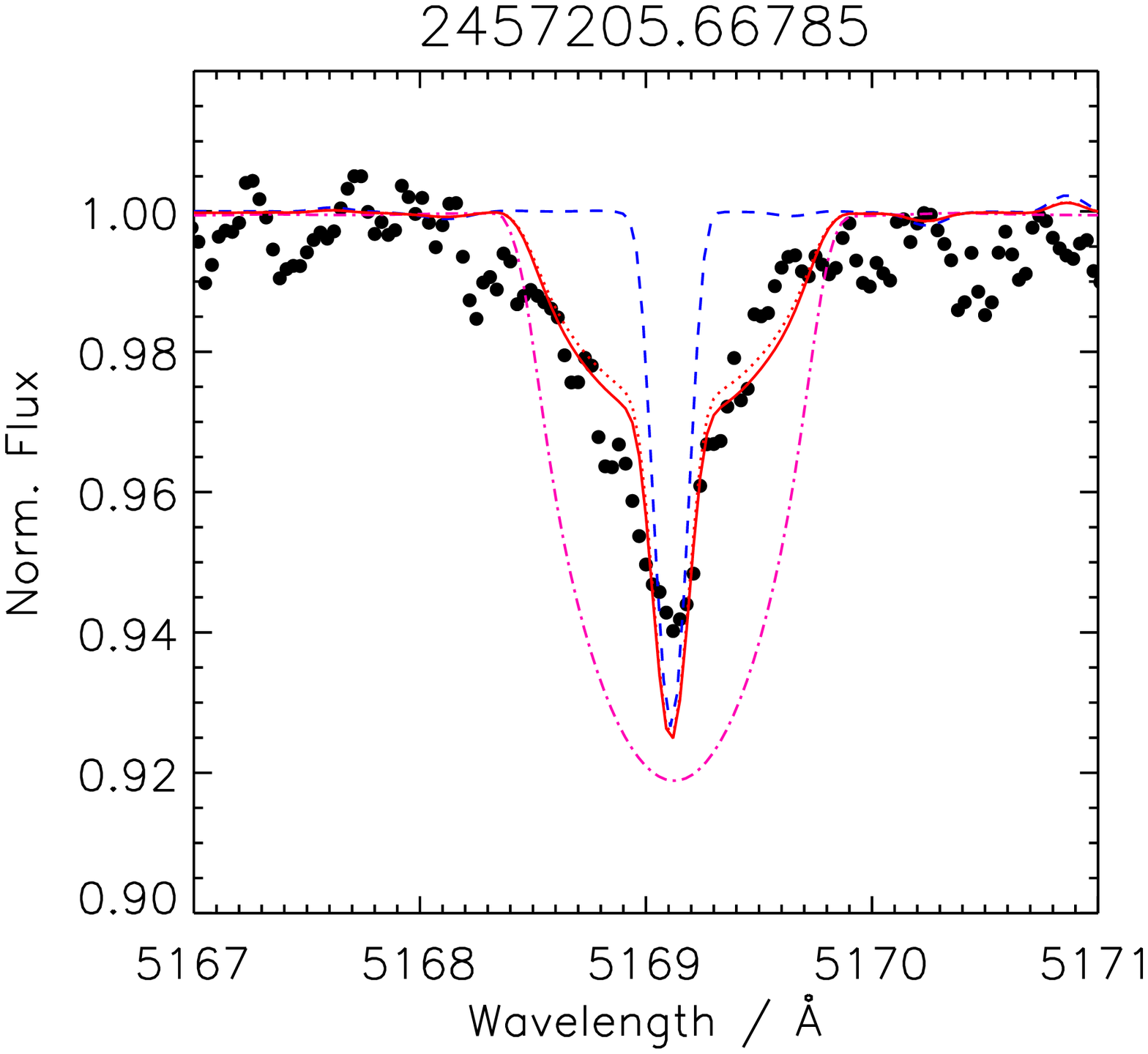} &
   \includegraphics[width=.225\textwidth]{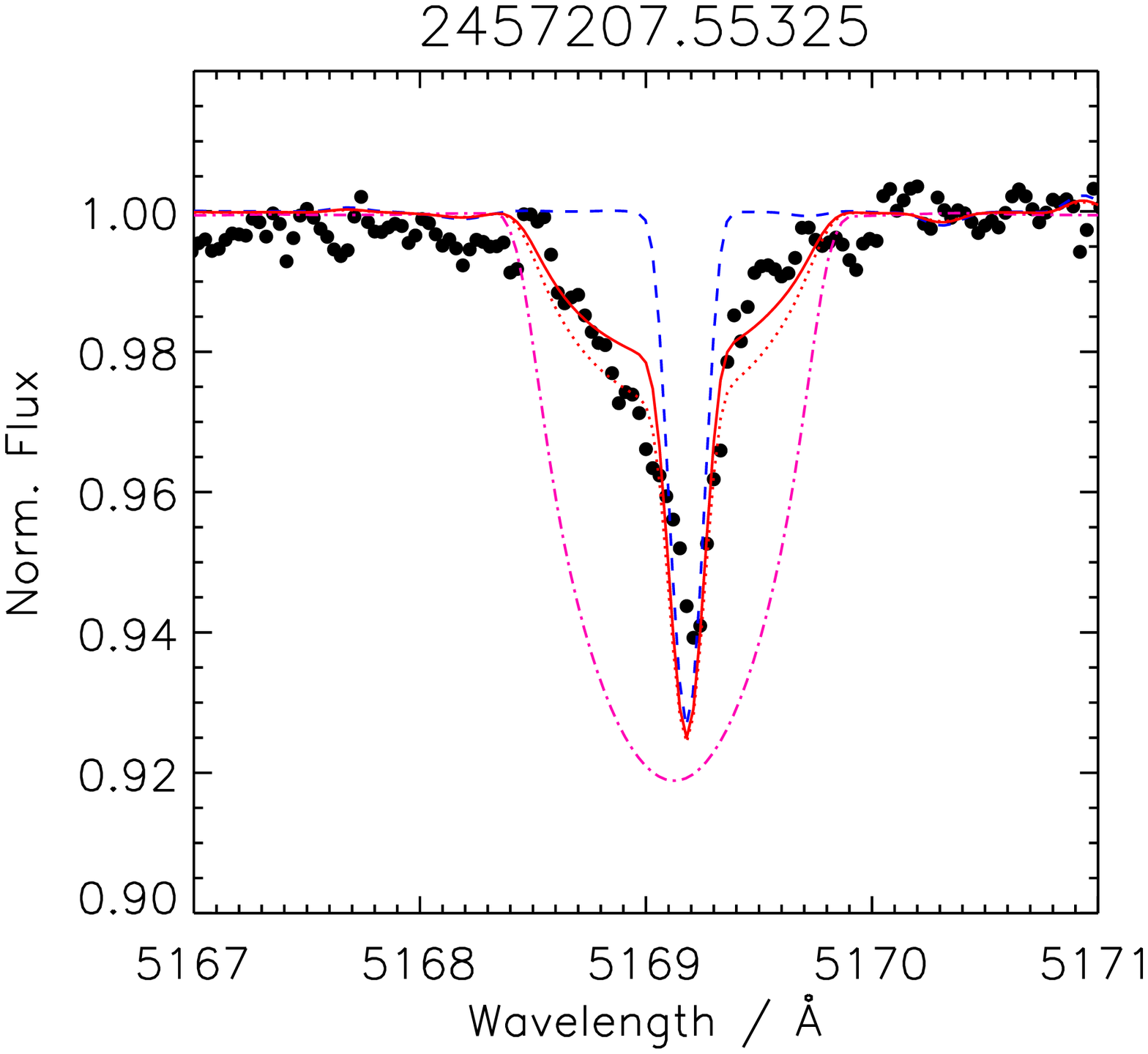} \\
   \includegraphics[width=.225\textwidth]{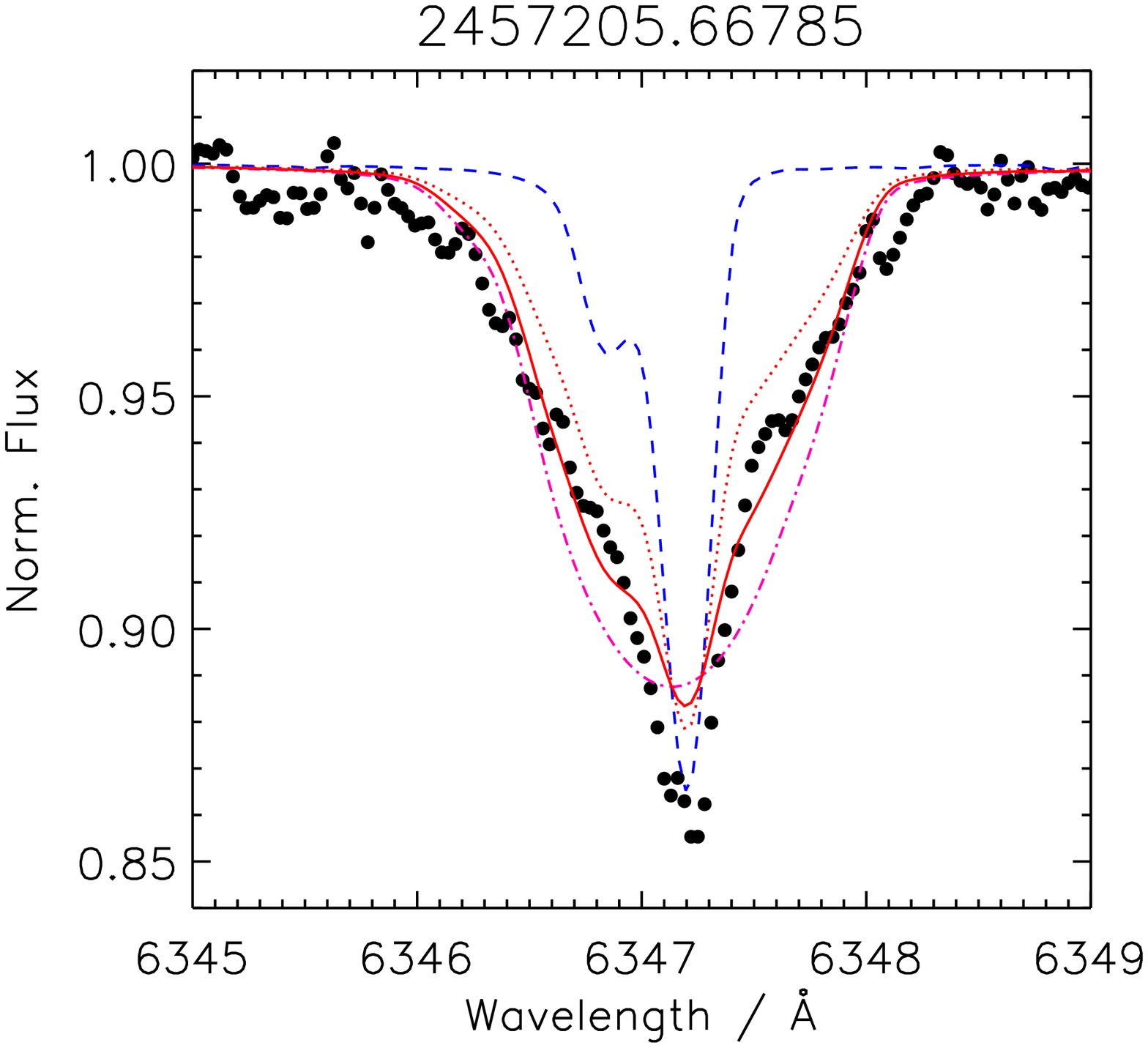} &
   \includegraphics[width=.225\textwidth]{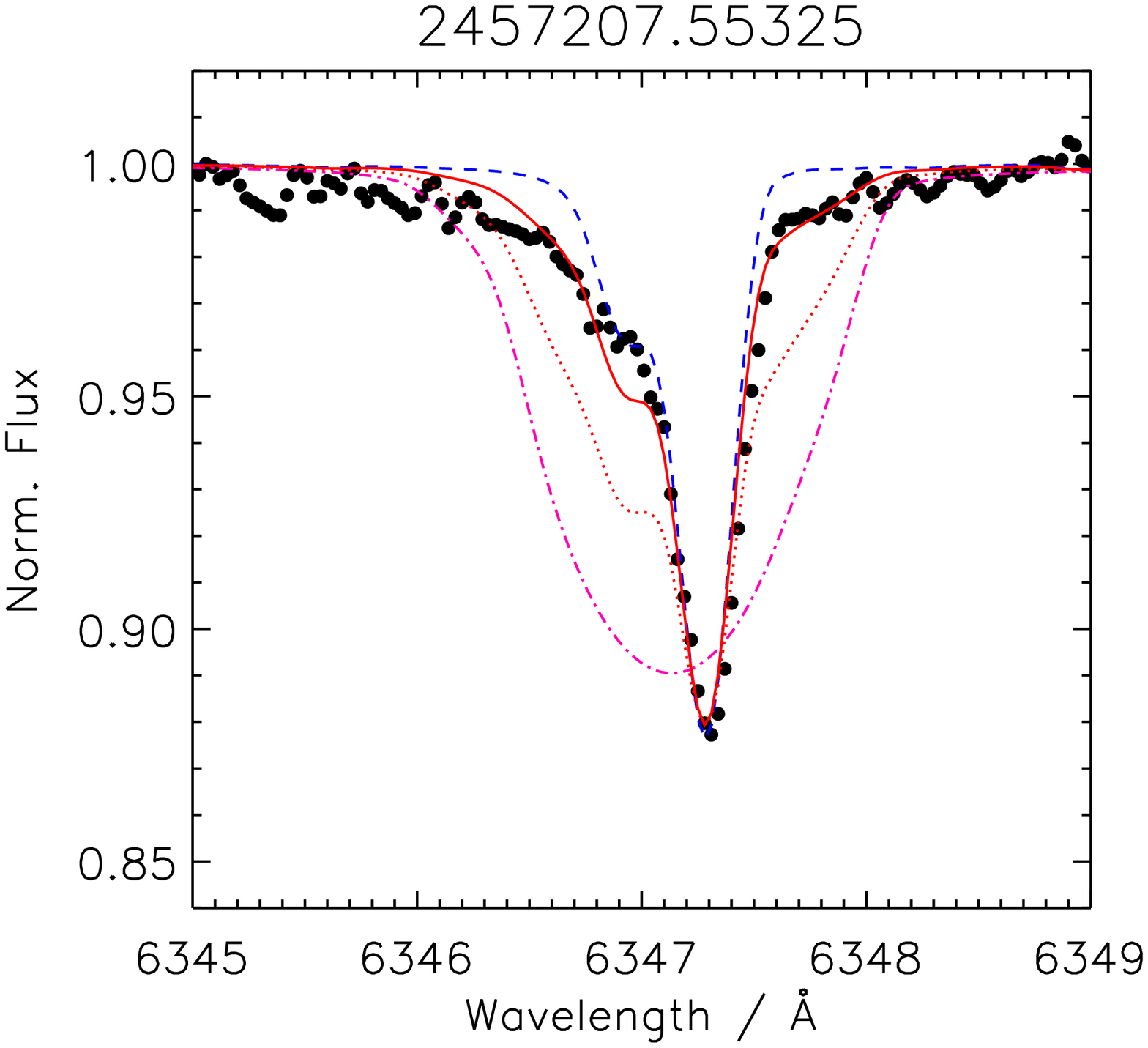} \\
   \includegraphics[width=.225\textwidth]{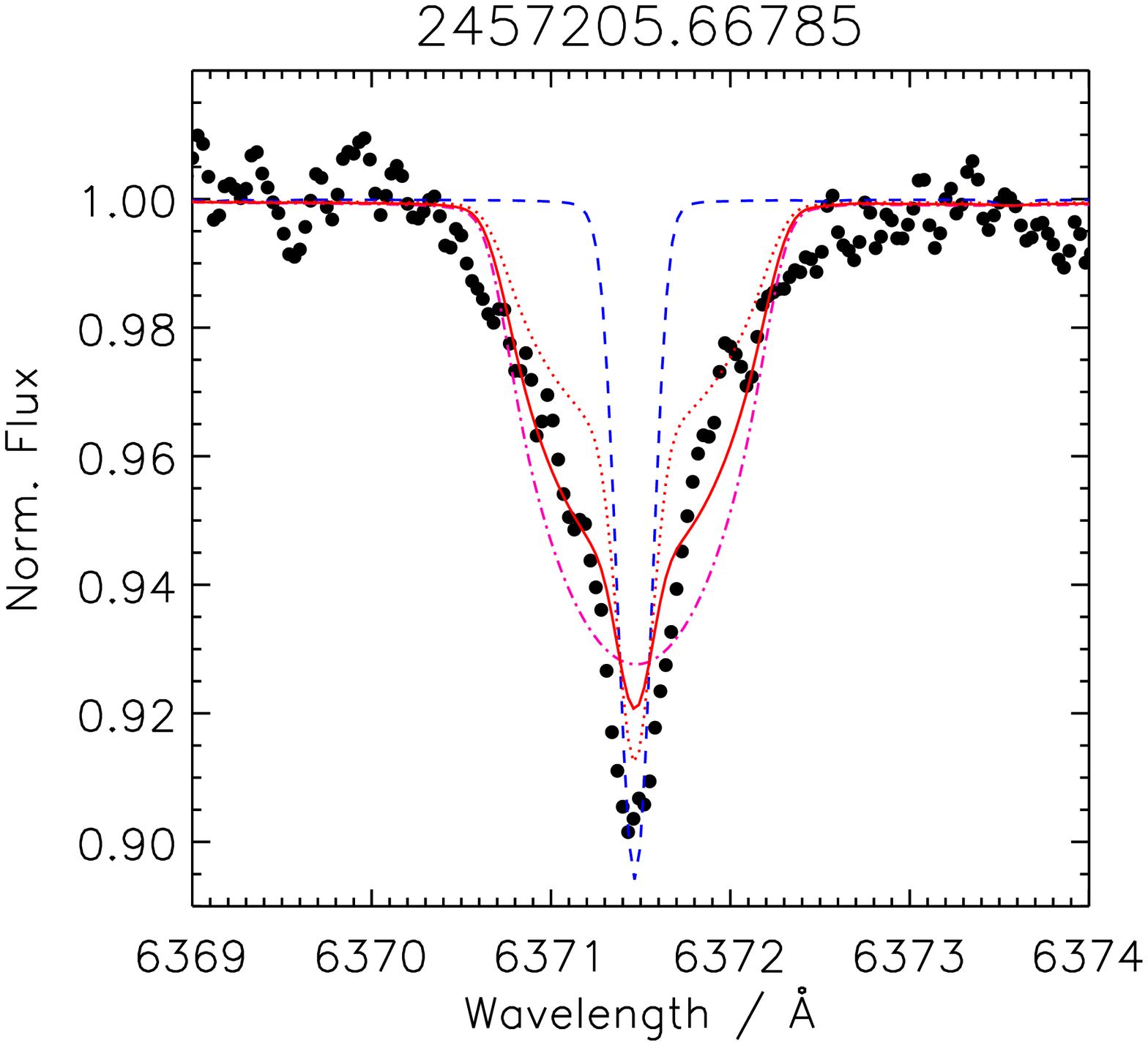} &
   \includegraphics[width=.225\textwidth]{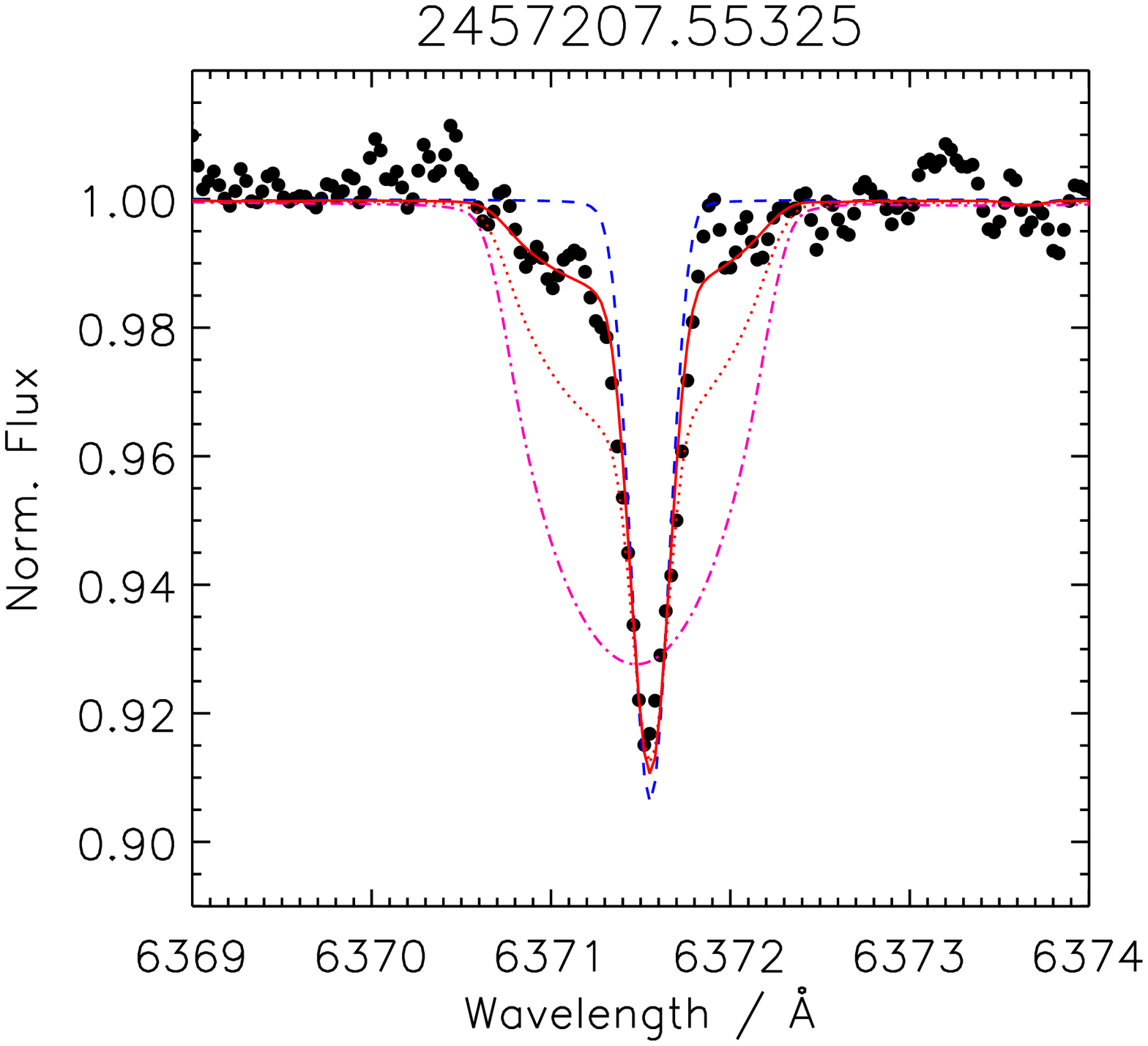} \\
   \includegraphics[width=.225\textwidth]{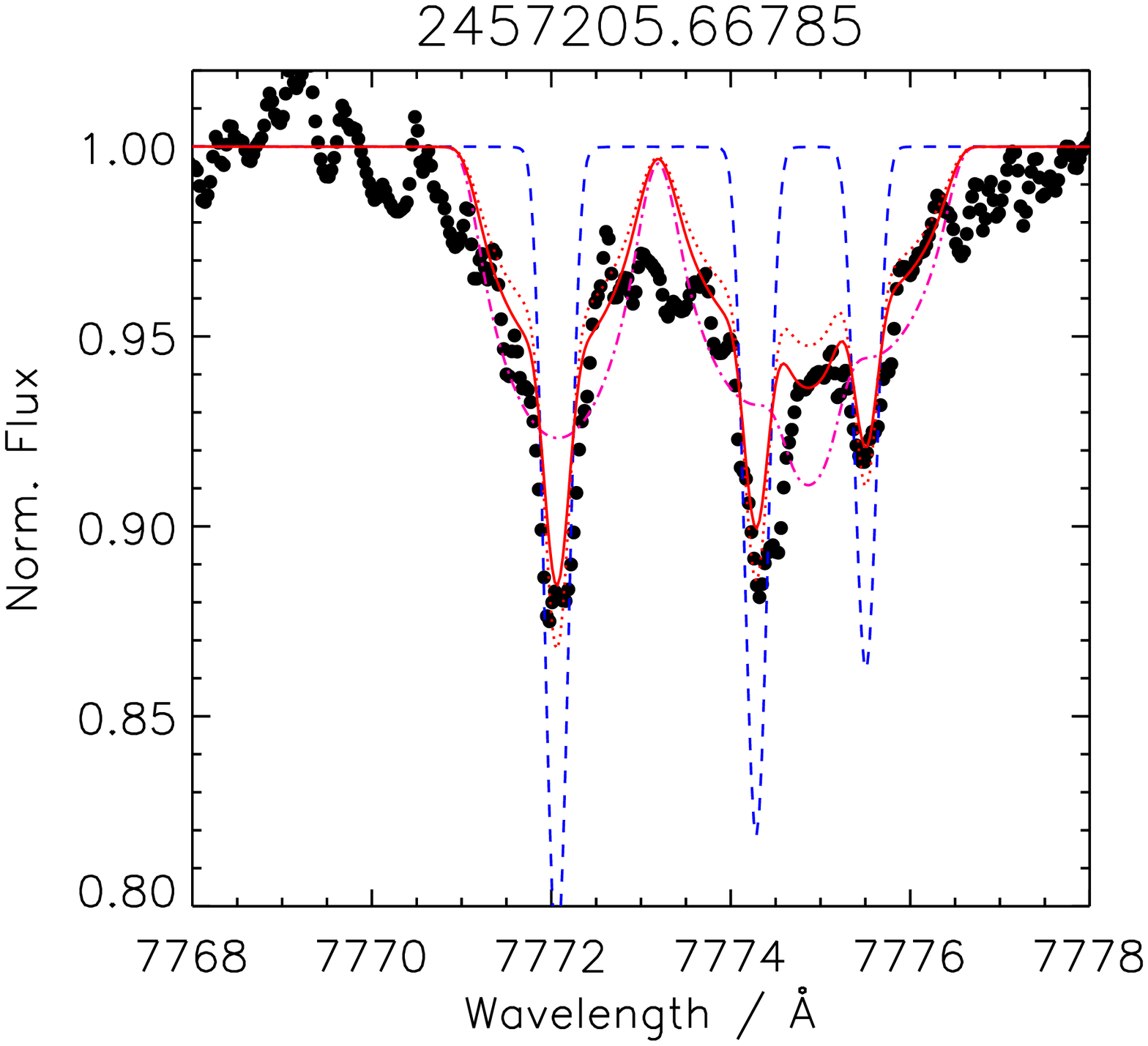} &
   \includegraphics[width=.225\textwidth]{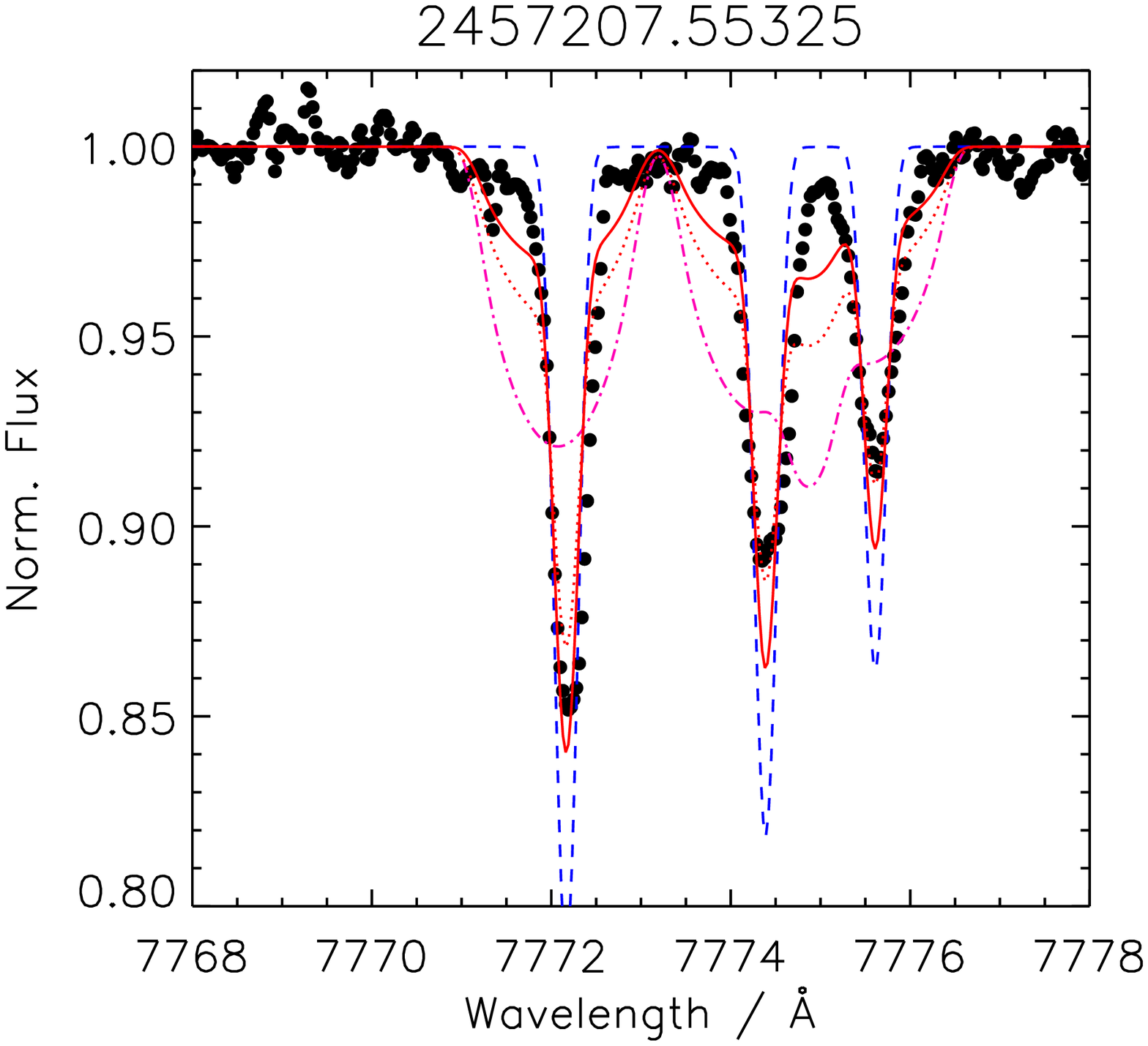} \\
\end{tabular}
      \caption[]{{\em Top to bottom}: Fits to the Fe~{\sc ii} $\lambda\lambda$ 5169, Si~{\sc ii} $\lambda\lambda$ 6347, and Si~{\sc ii} $\lambda\lambda$ 6371 lines, and the O~{\sc i} $\lambda\lambda$ 7774 triplet. Black circles: observations; dashed blue: primary; dot-dashed purple: secondary; dotted red: overall best-fit to all spectra; solid red: best fit to the individual spectrum. Note the variable strength of the contribution from the secondary.}
         \label{vsini}
   \end{figure}

   \begin{figure}
   \centering
   \includegraphics[width=8.5cm]{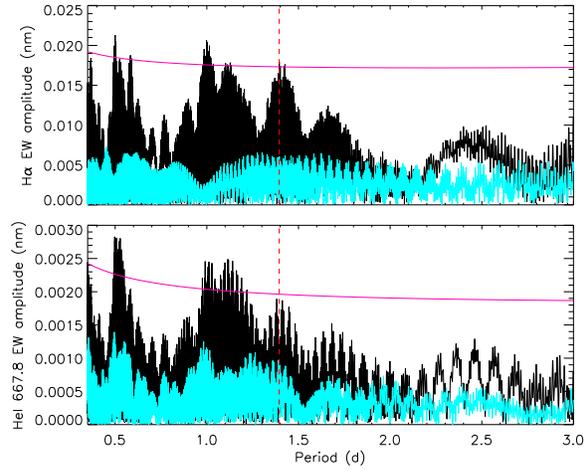} 
      \caption[]{Periodograms for H$\alpha$ emission EWs (top) and He~{\sc i}~6678 EWs (bottom). Light blue shows periodograms after pre-whitening with the 0.52~d period. The dashed red line shows the 1.4~d period determined from TESS data. The purple curve shows 3$\times$ the noise level.}
         \label{halpha_periods}
   \end{figure}

   \begin{figure}
   \centering
   \includegraphics[width=0.45\textwidth]{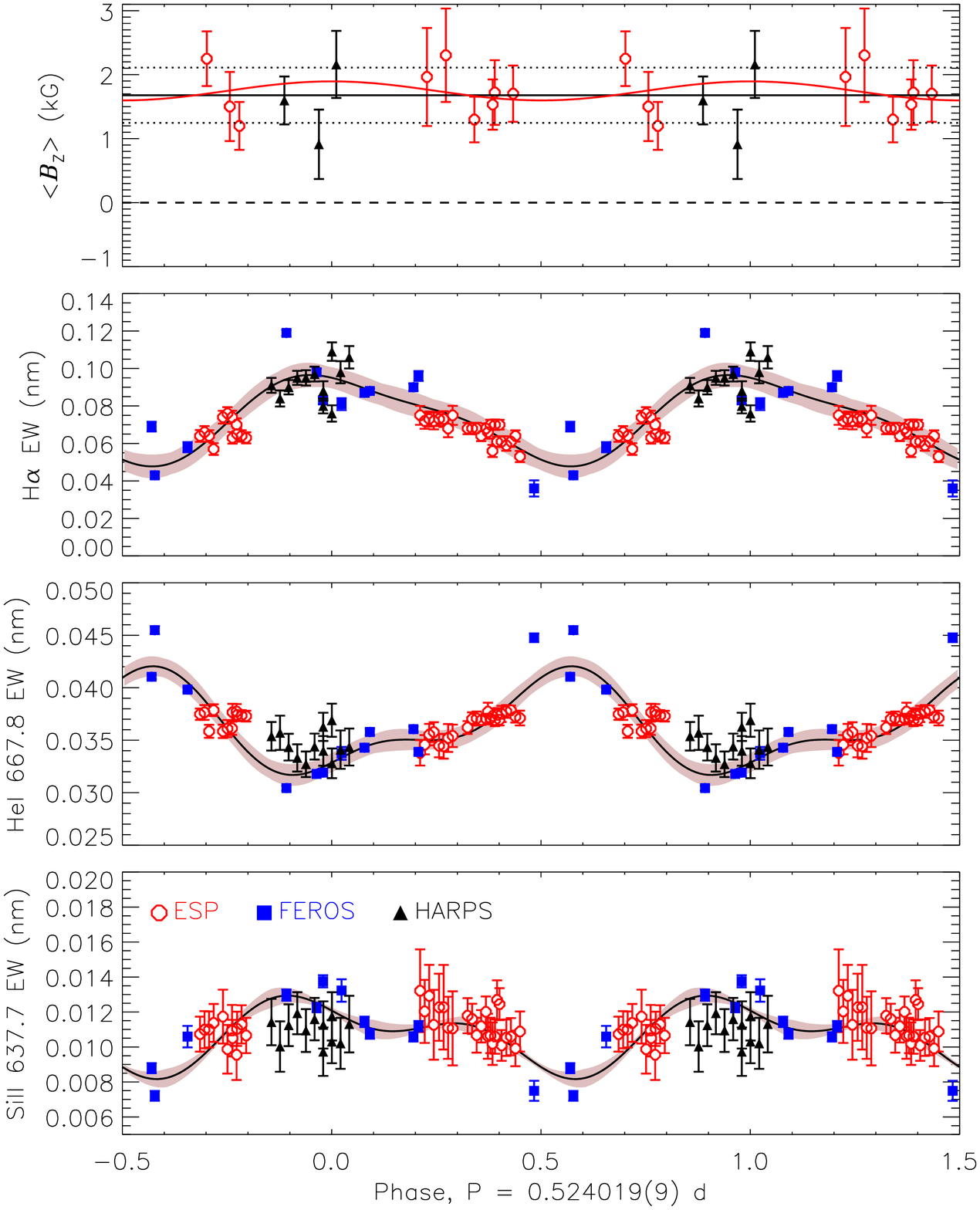} 
      \caption[]{{\em Top to bottom}: \bz, H$\alpha$ EW, He~{\sc i} 6678 EW, and Si~{\sc ii} 6347 EW, phased with the rotation period of HD\,156424B. EW curves show harmonic fits; shaded regions show fit uncertainties. In the top panel, the solid and dotted lines indicate the mean \bz, and the standard deviation, which is comparable to $\sigma_B$. The red curve shows the inferred magnetic model.}
         \label{halpha_ew}
   \end{figure}

As can be seen from Fig.\ \ref{hd156424_halpha}, and as was reported by \cite{alecian2014}, HD\,156424 displays H$\alpha$ emission characteristic of a centrifugal magnetosphere \citep[CM;][]{petit2013}: two emission bumps at high velocities, presumably originating from two magnetically confined clouds within the warped disk of the CM \citep[e.g.][]{lb1978,town2005c}. HD\,156424 is however an anomalous case. In general, the emission should be outside of the Kepler corotation radius \rk; since the magnetically confined plasma is in strict corotation with the star, there is a linear mapping between line-of-sight velocity and projected distance. Assuming that the emission belongs to the primary, the emission peaks at about $\pm$300~\kms~would then correspond to a projected distance of about 35 stellar radii. While this is outside the value of \rk~$=4.2~R_*$ determined by \cite{2019MNRAS.490..274S}, it is also outside the Alfv\'en radius \ra~$=21~R_*$, which is clearly impossible. While these parameters change given the stronger \bz~and higher \teff~inferred here for HD\,156424, \ra~is very unlikely to become much larger since the higher \mdot~will compensate for an increase in $B_{\rm d}$ (a lower limit of 12~$R_*$ is inferred in \S~\ref{sec:magpars}).

If, on the other hand, the emission belongs to the previously unrecognized secondary, then the emission peaks occur at a somewhat more reasonable $11~R_*$; while higher than the 3 or 4 $R_*$ at which CM emission peaks are generally observed, this is not unheard of, as in the case of CPD\,$-62^\circ 2124$, for which \cite{2017A&A...597L...6C} found the emission peak to occur at about 8~$R_*$. HD\,156424B is apparently rotating more rapidly than HD\,156424A, and has a magnetic field at least twice as strong, with a minimum surface dipole strength of about 5 kG. \cite{2019MNRAS.490..274S} showed that stars with CM-type H$\alpha$ emission are exclusively rapid rotators with very strong magnetic fields, making the secondary an inherently better candidate as the host of the emission. 

Assuming that the H$\alpha$ emission is formed in the CM, it should be modulated purely by rotation. H$\alpha$ equivalent widths were measured in an attempt to determine the rotational period. For the ESPaDOnS and FEROS data, un-normalized spectra were utilized, with a simple linear normalization performed on either side of the integration range; this was done to avoid warping of the line profiles due to the usual polynomial normalization process. Unnormalized spectra are not available for HARPSpol. In order to minimize scatter due to the RV variation of HD\,156424, synthetic H$\alpha$ spectra were calculated incorporating the flux of both components, assuming \teff$_{\rm A} = 23$~kK, \teff$_{\rm B} = 16$~kK, $\log{g} = 4.0$ in both cases, and shifting the individual spectra to their respective RVs. These fits are shown for maximum and minimum emission in the bottom panel of Fig.\ \ref{hd156424_halpha}.

   \begin{figure*}
   \centering
\begin{tabular}{ccc}
   \includegraphics[width=0.45\textwidth]{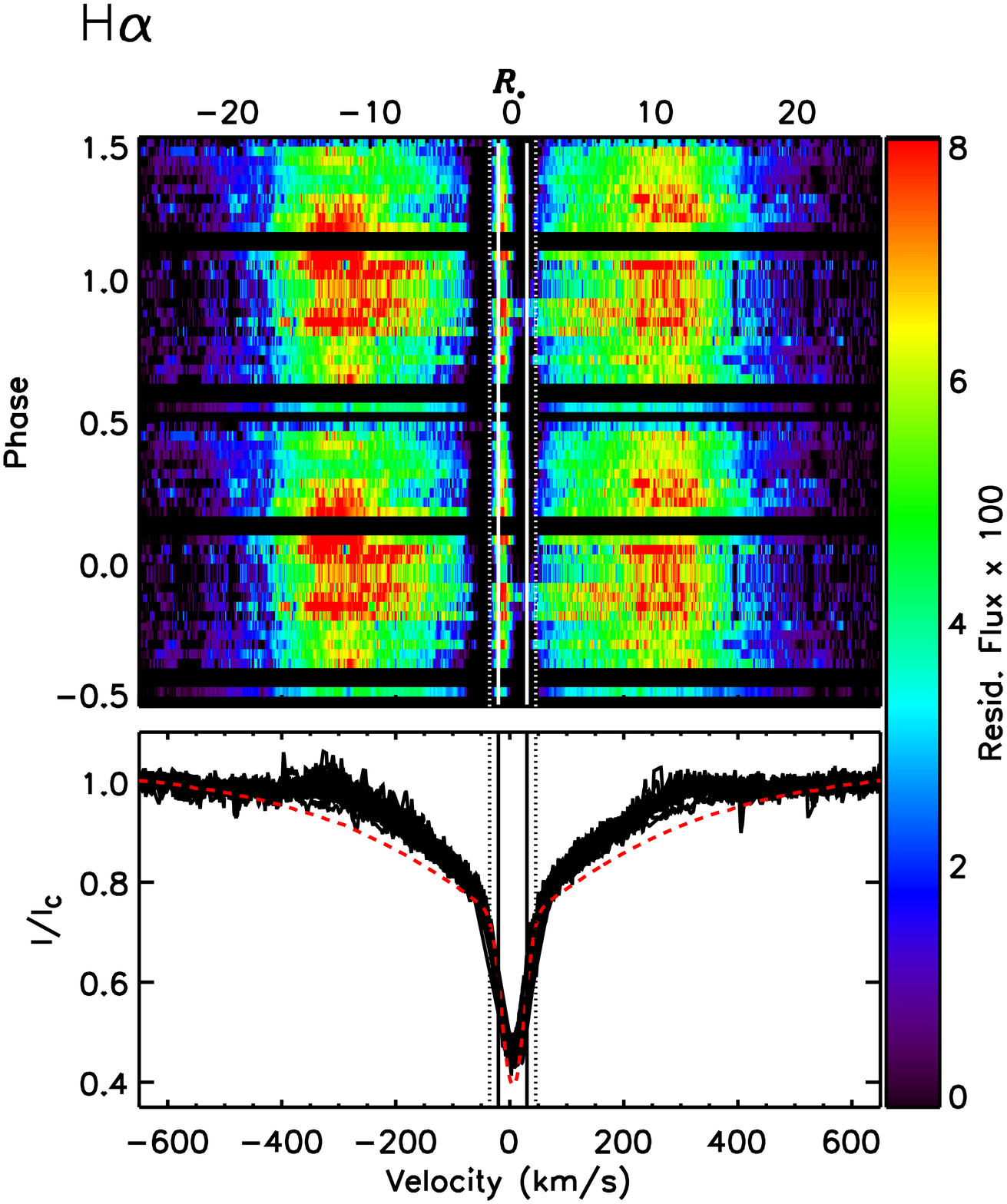} &
   \includegraphics[width=0.45\textwidth]{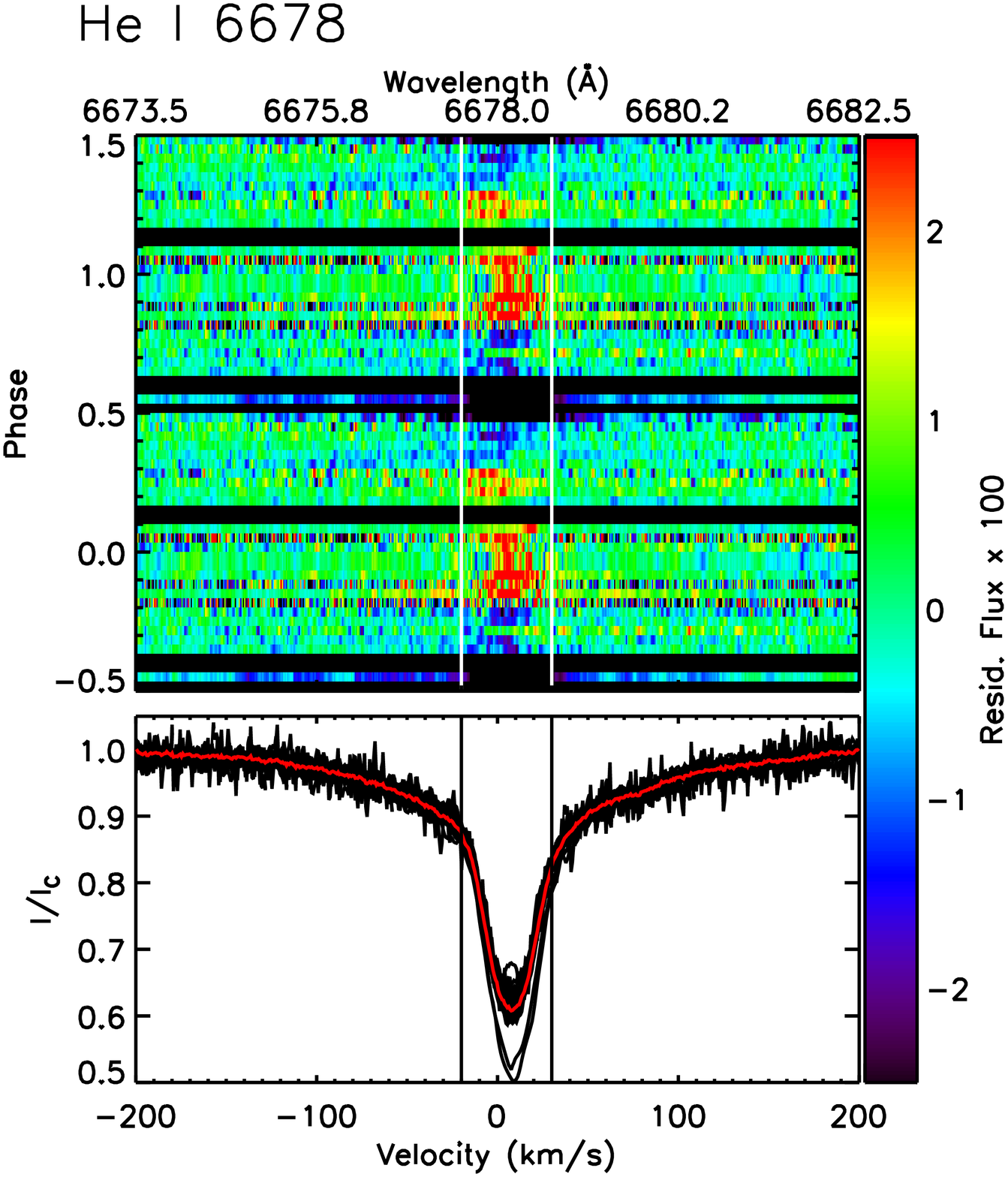} \\
\end{tabular}
      \caption[]{Dynamic spectra of H$\alpha$ and He~{\sc i} 6678. The top horizontal axis of the H$\alpha$ dynamic spectrum converts radial velocity to stellar radius under the assumption that line formation happens within a corotating magnetosphere. For H$\alpha$, individual synthetic spectra were used to obtain the residual flux, with a representative synthetic spectrum shown in the bottom panel by the dashed red line; for He~{\sc i}~6678, a mean spectrum was used, shown in the bottom panel by a solid red line. Vertical solid lines in both panels indicate $\pm$\vsini; for H$\alpha$, \rk~is indicated with vertical dotted lines.}
 
        \label{halpha_dyn}
   \end{figure*}

EWs were measured in the red and blue wings of the residual flux profiles, i.e.\ outside of $\pm$\vsini, with the outer limits defined by the maximum velocity of emission (i.e.\ about 500~\kms; dotted green lines in Fig.\ \ref{hd156424_halpha}). These emission EWs were analyzed individually and combined; the periodogram for the total EWs is shown in Fig.\ \ref{halpha_periods}. Maximum power is found at 0.52403(1)~d for blue, 0.52681(2)~d for red, and 0.524019(9)~d for the total EWs. 

Fig.\ \ref{vsini} shows synthetic spectral TLUSTY fits to Fe~{\sc ii} $\lambda\lambda$ 5169, Si~{\sc ii} $\lambda\lambda$ 6347, Si~{\sc ii} $\lambda\lambda$ 6371, and the O~{\sc i} $\lambda\lambda$ 7774 triplet, all lines in which the contribution of the secondary is relatively strong. The strength of the secondary's contribution is variable: the left column shows the FEROS observation in which HD\,156424B's contribution is at a maximum, the right column a FEROS observation in which it is at a minimum. O and Si are more variable than Fe. This is probably due to the presence of chemical abundance patches on the secondary. Under the assumption that these line profile variations are due to spots, they should also show rotational modulation. We additionally measured the EWs of He~{\sc i} 6678 and Si~{\sc ii}~6347, which are both relatively strong, and appear in all 3 spectroscopic datasets. Period analysis of the EWs measured from these lines finds maximum amplitude at 0.52399(1)~d for He~{\sc i}~6678 (Fig.\ \ref{halpha_periods}, bottom), and 0.53713(5)~d for Si~{\sc ii}~6347, essentially consistent with the results from H$\alpha$.

Notably, the 1.39~d period obtained from the TESS photometry does not appear in any of the EW datasets (red dashed lines in Fig.\ \ref{halpha_periods}). It therefore seems likely that the TESS period is not due to rotation, but may instead be due to gravity-mode pulsations in HD\,156424A. Similarly, the 0.52~d period from the EWs does not appear in the TESS light curve. This may be a consequence of the light curve being dominated by the variability of HD\,156424A, with the contribution of the dimmer HD\,156424B only appearing in individual spectral lines in which its strong surface abundance patches lead to EW variations. Typical mCP stars have light curve variations on the order of 1 to 10 mmag \citep[e.g.][]{2019MNRAS.487..304D,2019MNRAS.487.4695S}, or about 0.1\% to 1\% of flux. Assuming HD\,156424A is about 8$\times$ brighter than HD\,156424B, the flux modulation due to chemical spots would decrease to about 0.0125\% to 0.125\%, or around 0.15 to 1.5 mmag. The pre-whitened TESS light curve has an upper limit of around 0.15 mmag near 2 c/d, making it possible that the contribution from HD\,156424B's rotational modulation remains hidden in the noise. Note that the star's relatively low \vsini~and very rapid rotation indicate that the rotational axis must be nearly aligned with the line of sight, in which case the photometric variation due to chemical spots should be very low in amplitude. 


The EWs are shown phased with the 0.52~d period in Fig.\ \ref{halpha_ew}, using $T0 = 2456126.21(4)$ as determined from the maximum emission strength of a harmonic fit. Dynamic spectra of H$\alpha$ and He~{\sc i}~6678 are shown in Fig.\ \ref{halpha_dyn}. Emission is detectable at all phases, with only weak rotational modulation. This suggests that the obliquity of the magnetic field must be relatively small, with magnetically confined plasma almost equally distributed within the magnetic equatorial plane \citep[e.g.][]{town2005c}. The low level of variability in H$\alpha$ also suggests that the light curve modulation due to rotational modulation of HD\,156424B is likely to be fairly modest, consistent with the failure to detect it in the TESS data.

The \bz~measurements for HD\,156424B do not phase coherently with the 0.52~d period determined from H$\alpha$ (Fig.\ \ref{halpha_ew}, top); however, this is not surprising given that 1) the mean error bar is larger than the standard deviation in \bz; and 2) it is probable that the \bz~measurements of both components are contaminated by the Stokes $V$ contribution from the other component. 

\section{Magnetic Models}\label{sec:magpars}

\begin{table}
\centering
\caption[]{Rotational, magnetic, and magnetospheric parameters. Values for $B_{\rm d}$ for HD\,156424A correspond to the lower limit and the 3$\sigma$ upper limit.}
\label{magtab}
\begin{tabular}{l c c}
\hline\hline
Parameter & A  & B\\
\hline
$R_{\rm p}~({\rm R}_\odot)$ & $3.8 \pm 0.2$ & $2.5 \pm 0.1$ \\
$M_*~({\rm M}_\odot)$ & $8.8 \pm 0.6$ & $4.8 \pm 0.4$ \\
$P_{\rm rot}~({\rm d})$ & -- &  0.524019(9) \\
$T_0~({\rm HJD})$ & -- & 2456126.22(4) \\
\vsini~(\kms) & $5 \pm 1$ & $25 \pm 2$ \\
$v_{\rm mac}$~(\kms) & $14 \pm 1$ & -- \\
$i_{\rm rot}~(^\circ)$ & $<20$ & $6 \pm 1$ \\
$v_{\rm eq}~({\rm kms})$ & $>6$ & $270^{+15}_{-8}$ \\
$W$ & $>0.01$ & $0.47^{+0.03}_{-0.01}$ \\
$R_{\rm p}/R_{\rm e}$ & $1$ & $0.913^{+0.003}_{-0.01}$ \\
$R_{\rm K}~({\rm R_*})$ & $<10$ & $1.65 \pm 0.04$ \\
\hline
$B_0~({\rm kG})$ & $0.82 \pm 0.05$ & $1.6 \pm 0.1$ \\
$B_1~({\rm kG})$ & $0.06 \pm 0.05$ & $0.4 \pm 0.6$ \\
$\beta~(^\circ)$ & $<20$ & $31^{+20}_{-31}$ \\
$B_{\rm d}~({\rm kG})$ & $>3/<16.2$ & $8^{+5}_{-2}$ \\
\hline
$\log{(\dot{M} / {\rm M_\odot / yr})}$ & $-9.1^{+0.2}_{-0.4}$ & $-10.7 \pm 0.2$ \\
$v_\infty~({\rm km~s^{-1}})$ & $1100^{+600}_{-50}$ & $1100^{+10}_{-13}$ \\
$\log{\eta_*}$ & $4.3 \pm 0.3$ & $6.8 \pm 0.5$ \\
$R_{\rm A}~({\rm R_*})$ & $>12$ & $43^{+8}_{-3}$ \\
$\log{R_{\rm A}/R_{\rm K}}$ & $<0.4$ & $1.43 \pm 0.05$ \\
$\log{(B_{\rm K} / {\rm G})}$ & $<0.8$&  $3.0 \pm 0.1$ \\
\hline\hline
\end{tabular}
\end{table}

\subsection{HD\,156424A}

Since there is no sign of line profile variability coherent with the 1.39~d period identified in TESS photometry, this period probably does not reflect rotational modulation. It is furthermore doubtful that the \bz~measurements can be relied upon to determine the rotational period, as they likely remain contaminated by polarization from HD\,156424B. Therefore the rotational period of this star cannot be determined. 

Examination of the Stokes $V$ profiles shows that they are not strongly variable, in all cases presenting a typical Zeeman s-curve. This indicates that either $i$ or $\beta$, or more probably both, are small. First, the low \vsini~suggests either small $i$, very slow rotation, or both. There is no indication of long-term evolution over the $\sim$2 years of the spectropolarimetric dataset, arguing against slow rotation (unless the rotation period is exceptionally long, i.e.\ several decades). Second, a small $i$ and large $\beta$, or large $i$ and small $\beta$, would tend to produce cross-over signatures, which are not detected. 

HD\,156424A's magnetic field and magnetospheric parameters were therefore determined under the conservative assumption that $i < 20^\circ$, chosen to be small enough to reflect something close to the true value, yet large enough to permit reasonable coverage of the angular parameter space. The Monte Carlo Hertzsprung-Russell Diagram sampler described by \cite{2019MNRAS.490..274S} was utilized, with the parameter space constrained by the star's membership in the Sco OB4 association \citep[main-sequence turnoff age $\log{t/{\rm yr}} = 6.8 \pm 0.2$;][]{2005AA...438.1163K}. Since a fit could not be performed to \bz~without a rotation period, the vertical offset and semi-amplitude of the sinusoidal fit were approximated with the mean and standard deviation of \bz. This results in $\beta < 20^\circ$ and $B_{\rm d} > 3$~kG, with 2 and 3$\sigma$ upper limits of 6.2 kG and 16.2 kG.

The Alfv\'en radius \ra~was calculated in the same fashion as by \cite{2019MNRAS.490..274S}, i.e.\ using the scaling relationships provided by \cite{ud2002,ud2008}, and with the mass-loss rate \mdot~and wind terminal velocity \vinf~calculated using the \cite{vink2001} recipe. The upper limit on the Kepler corotation radius \rk~and the lower limit on \ra~suggest that the star probably has a small CM ($\log{R_{\rm A}/R_{\rm K}} < 0.4$), but the strength of the magnetic field at the Kepler radius $\log{B_{\rm K} / {\rm G}} < 0.8$ is much lower than the threshold for emission determined empirically by \cite{2020arXiv200912336S}. 

While the lack of a rotation period means that only a lower limit can be determined for \ra$> 12 R_*$, given the low \vsini~of HD\,156424A this lower limit translates to a projected velocity of 60~\kms, almost 10$\times$ less than the maximum extent of H$\alpha$ emission (Fig.\ \ref{hd156424_halpha}). HD\,156424A would need to have \ra$> 100 R_*$ in order for the H$\alpha$ emission to plausibly belong to this star. Since \ra~scales approximately as $B_{\rm d}^{1/2}$ \citep{ud2002,ud2008}, this would require $B_{\rm d} > 200$~kG, almost 10$\times$ stronger than the strongest magnetic field seen in an early-type star \citep{2019MNRAS.490..274S}. Indeed, such a strong magnetic field would produce obvious Zeeman splitting in the star's spectral lines, which is not observed. It can therefore be firmly excluded that the H$\alpha$ emission originates in a CM around HD\,156424A.

\subsection{HD\,156424B}

We proceed on the assumption that the 0.52~d period identified from H$\alpha$ is the rotational period of the secondary. The parameter space was constrained with the main-sequence turnoff age of Sco OB4. Since a fit to \bz~cannot be obtained, as with HD\,156424A we initially approximated the fitting parameters $B_0$ and $B_1$ \citep[see][]{2018MNRAS.475.5144S} from the mean and standard deviation of \bz, respectively. However, the large standard deviation in \bz, resulting in a geometrical parameter $r = \cos{(i+\beta)}/\cos{(i-\beta)} = (B_0 - B_1)/(B_0 + B_1) = 0.6 \pm 0.5$ \citep{preston1967,preston1974}, resulted in a maximum likeliehood value of $\beta$ that did not give a good reproduction of the actual measurements. The standard deviation in \bz~is less than the mean error bar, and therefore probably over-estimates the actual underlying variation in \bz. We instead adopted the mean weighted uncertainty in \bz, 0.1 kG, as both the semi-amplitude $B_1$ and the uncertainty in the semi-amplitude, yielding $r = 0.88 \pm 0.12$. The resulting dipolar ORM model is compared to the \bz~measurements in the top panel of Fig.\ \ref{halpha_ew}.

The result is that the inclination $i \sim 6^\circ$ is very small, as is $\beta \sim 30^\circ$, although with a large error bar. The very small $i$ is consistent with the relatively low level of variation in the EWs and \bz. The small $\beta$ is likewise consistent with the minimal variation in the LSD Stokes $V$ profiles. The surface dipole strength $B_{\rm d} \sim 8$~kG. The very rapid rotation means that the star must have a non-negligible oblateness ($R_{\rm p}/R_{e} \sim 0.91$), and a very small Kepler radius (1.6~$R_*$). This in turn means that $\log{B_{\rm K} / {\rm G}} = 3.0$, which is close to the upper extreme of the sample examined by \cite{2020arXiv200912336S}. 

The peak H$\alpha$ emission strength is about 0.08 nm. In the vicinity of H$\alpha$, the synthetic BSTAR2006 spectra used to measure the EWs indicate that HD\,156424B should be about 20\% as bright as the primary, thus the EW should be scaled by a factor of 5 for an intrinsic emission strength of 0.4 nm. Comparing this to Fig.\ 5 by \cite{2020arXiv200912336S}, this is almost exactly the predicted emission strength for a star with HD\,156424A's extreme value of $B_{\rm K}$. The analytical scaling relationship for H$\alpha$ emission EW from CMs based on centrifugal breakout, given by \cite{2020arXiv200912359O}, yields $0.36 \pm 0.08$~nm, again corresponding very closely to the measured value once corrected for dilution by the light of the primary.

While the emission strength is consistent with $B_{\rm K}$, the extent of emission is quite surprising. In all of the stars examined by \cite{2020arXiv200912336S}, the radius at which emission is at a maximum is between 1 and 2~$\times$~\rk. In this case, assuming a linear mapping between radius $r$ and velocity $v$ such that $r/R_* = v/v\sin{i}$, and with \vsini~$= 25$~\kms, the emission peak occurs at about $r_{\rm max} \sim 10~R_*$, which is several times larger than $R_{\rm K}$. The extent of emission is also extraordinary: about 20~$R_*$, as compared to the next-most-extensive magnetospheric emission profile, that of CPD$-62^\circ 2124$, about 15~$R_*$. For $r_{\rm max}$ to occur between 1 and 2$\times$\rk, \vsini~would need to be between about 100 and 200~\kms. Such high values of \vsini~are clearly excluded by the data. If the rotation period is actually twice as long, i.e. 1.04~d, the Kepler radius is about 2.8~$R_*$, and the emission peak then occurs at about 4.4~$R_{\rm K}$ -- still siginficantly further from \rk~than is generally seen. 

\section{Discussion and Conclusions}\label{sec:discussion}

We have analyzed the TESS data and the combined high-resolution spectropolarimetric and spectroscopic dataset for HD\,156424. Several frequencies are detectable in the TESS light curve, 4 with amplitudes of 2 to 3 mmag, the remainder with amplitudes of order 0.1 mmag; 6 of the 11 detected frequencies are independent, the remainder being harmonics or linear combinations. The majority of the frequencies are above 5 c/d, indicating that HD\,156424 is a $\beta$ Cep pulsator, although the second-strongest frequency at about 0.72~${\rm d}^{-1}$ could be indicative of Slowly Pulsating B-star (SPB) pulsation. Analysis of RVs also detects the strongest frequency, confirming that these oscillations belong to HD\,156424 and not a background star. RVs also show evidence for long-term RV variations with an amplitude of $\sim 5$~\kms~and a period of order years. The small but statistically significant difference in frequencies obtained from the TESS and RV datasets (separated by several years) is consistent with the light-time effect from orbital motion with the observed RV amplitude. 

Close examination of the mean spectrum has revealed the presence of a companion star, HD\,156424B. This star is not a RV variable and is therefore not the companion responsible for HD\,156424A's orbital RV variation; instead, we identify it as the orbital companion detected via speckle observations by \cite{1993AJ....106..352H,2010AJ....139..743T}. In an attempt to remove the contribution of HD\,156424B's spectrum from the LSD mean line profile and thereby obtain cleaner magnetic measurements, we found that HD\,156424B is itself a magnetic star. This makes HD\,156424 the second known doubly magnetic hot binary although, unlike $\epsilon$ Lupi, the stars are not interacting \citep{2015MNRAS.454L...1S,2019MNRAS.488...64P}. Neither HD\,156424A nor HD\,156424B have strongly variable Stokes $V$ profiles. The mean \bz~for HD\,156424A is about $-800$~G, while that of HD\,156424B is about $+1.5$~kG. In neither case could a rotational period be determined from \bz, likely due to a combination of residual contamination of Stokes $V$ by the other star, combined with the overall low level of variation. 

The spectrum displays variable H$\alpha$ emission with a morphology consistent with an origin in a Centrifugal Magnetosphere; that is, with two emission peaks at velocities greater than \vsini. He~{\sc i}, O~{\sc i}, Si~{\sc ii}, and Fe~{\sc ii} lines are also variable. The variability in Si~{\sc ii} is clearly due to HD\,156424B. Period analysis of the EWs of these lines reveals in each case a periodicity at about 0.52~d. Since this is likely the rotation period, HD\,156424B is apparently one of the most rapidly rotating magnetic B-type stars found to date. This 0.52~d periodicity does not appear in the TESS light curve, probably because HD\,156424A is about 2--8$\times$ brighter than HD\,156424B (the approximate range consistent with photometric and spectroscopic constraints), while the actual amplitude of the rotational variability in the latter case is low due to the system's small rotational inclination. There is no indication of the $0.72~{\rm d}^{-1}$ signal in the EWs, suggesting that this frequency is probably not due to rotation.

Adjusting for dilution of the spectrum by HD\,156424A, the H$\alpha$ emission is comparable to the strongest yet detected in a magnetic B-type star. This very strong emission is consistent with HD\,156424B's apparently rapid rotation and strong magnetic field. However, peak emission appears at about a distance of 10 $R_*$ from the star, very far above the Kepler radius. This is surprising, since in every other case emission strength peaks between $1$ and $2~R_{\rm K}$. The maximum extent of emission is furthermore about 20~$R_*$, indicating the star's magnetosphere is very extended compared to similar systems. 



\section*{Acknowledgements}

This work is based on observations obtained at the Canada-France-Hawaii Telescope (CFHT) which is operated by the National Research Council of Canada, the Institut National des Sciences de l'Univers (INSU) of the Centre National de la Recherche Scientifique (CNRS) of France, and the University of Hawaii; and at the La Silla Observatory, ESO Chile with the 3.6 m telescope and the MPA 2.2 m telescope under programme IDs and 187.D-0917(C), 095.D-0269(A), and 095.A-9007. This work has made use of the VALD database, operated at Uppsala University, the Institute of Astronomy RAS in Moscow, and the University of Vienna. This research has made use of the SIMBAD database, operated at CDS, Strasbourg, France. Some of the data presented in this paper were obtained from the Mikulski Archive for Space Telescopes (MAST). STScI is operated by the Association of Universities for Research in Astronomy, Inc., under NASA contract NAS5-2655. MES acknowledges support from the Annie Jump Cannon Fellowship, supported by the University of Delaware and endowed by the Mount Cuba Astronomical Observatory. ADU acknowledges support from the NSERC Postdoctoral Fellowship Program. GAW acknowledges support from the Natural Sciences and Engineering Research Council (NSERC) of Canada in the form of a Discovery Grant. OK acknowledges support by the Swedish Research Council and the Swedish National Space Agency. The MiMeS collaboration acknowledge financial support from the Programme National de Physique Stellaire (PNPS) of INSU/CNRS. We acknowledge the Canadian Astronomy Data Centre (CADC). The authors thank the anonymous referee for their helpful criticism. 

\section*{Data Availability Statement}
Reduced ESPaDOnS spectra are available at the CFHT archive maintained by the CADC at \url{https://www.cadc-ccda.hia-iha.nrc-cnrc.gc.ca/en/}, and FEROS and HARPSpol spectra are available in raw form at the ESO archive at \url{http://archive.eso.org/eso/eso_archive_main.html}. TESS data are available at the MAST archive at \url{https://mast.stsci.edu/portal/Mashup/Clients/Mast/Portal.html}. Data in all archives can be found via standard stellar designations. Reduced ESO data are available from the authors on request.

\bibliography{bib_dat.bib}{}

\begin{thebibliography}{}
\makeatletter
\relax
\def\mn@urlcharsother{\let\do\@makeother \do\$\do\&\do\#\do\^\do\_\do\%\do\~}
\def\mn@doi{\begingroup\mn@urlcharsother \@ifnextchar [ {\mn@doi@}
  {\mn@doi@[]}}
\def\mn@doi@[#1]#2{\def\@tempa{#1}\ifx\@tempa\@empty \href
  {http://dx.doi.org/#2} {doi:#2}\else \href {http://dx.doi.org/#2} {#1}\fi
  \endgroup}
\def\mn@eprint#1#2{\mn@eprint@#1:#2::\@nil}
\def\mn@eprint@arXiv#1{\href {http://arxiv.org/abs/#1} {{\tt arXiv:#1}}}
\def\mn@eprint@dblp#1{\href {http://dblp.uni-trier.de/rec/bibtex/#1.xml}
  {dblp:#1}}
\def\mn@eprint@#1:#2:#3:#4\@nil{\def\@tempa {#1}\def\@tempb {#2}\def\@tempc
  {#3}\ifx \@tempc \@empty \let \@tempc \@tempb \let \@tempb \@tempa \fi \ifx
  \@tempb \@empty \def\@tempb {arXiv}\fi \@ifundefined
  {mn@eprint@\@tempb}{\@tempb:\@tempc}{\expandafter \expandafter \csname
  mn@eprint@\@tempb\endcsname \expandafter{\@tempc}}}

\bibitem[\protect\citeauthoryear{{Alecian} et~al.,}{{Alecian}
  et~al.}{2014}]{alecian2014}
{Alecian} E.,  et~al., 2014, \mn@doi [\aap] {10.1051/0004-6361/201323286},
  \href {http://adsabs.harvard.edu/abs/2014A26A...567A..28A} {567, A28}

\bibitem[\protect\citeauthoryear{{Batten}}{{Batten}}{1973}]{1973bmss.book.....B}
{Batten} A.~H.,  1973, {Binary and multiple systems of stars}

\bibitem[\protect\citeauthoryear{{Bowman}, {Buysschaert}, {Neiner},
  {P{\'a}pics}, {Oksala}  \& {Aerts}}{{Bowman}
  et~al.}{2018}]{2018A&A...616A..77B}
{Bowman} D.~M.,  {Buysschaert} B.,  {Neiner} C.,  {P{\'a}pics} P.~I.,  {Oksala}
  M.~E.,   {Aerts} C.,  2018, \mn@doi [\aap] {10.1051/0004-6361/201833037},
  \href {https://ui.adsabs.harvard.edu/abs/2018A&A...616A..77B} {616, A77}

\bibitem[\protect\citeauthoryear{{Braithwaite}}{{Braithwaite}}{2009}]{2009MNRAS.397..763B}
{Braithwaite} J.,  2009, \mn@doi [\mnras] {10.1111/j.1365-2966.2008.14034.x},
  \href {http://cdsads.u-strasbg.fr/abs/2009MNRAS.397..763B} {397, 763}

\bibitem[\protect\citeauthoryear{{Braithwaite} \& {Spruit}}{{Braithwaite} \&
  {Spruit}}{2004}]{2004Natur.431..819B}
{Braithwaite} J.,  {Spruit} H.~C.,  2004, \mn@doi [\nat] {10.1038/nature02934},
  \href {http://cdsads.u-strasbg.fr/abs/2004Natur.431..819B} {431, 819}

\bibitem[\protect\citeauthoryear{{Breger} et~al.,}{{Breger}
  et~al.}{1993}]{1993A&A...271..482B}
{Breger} M.,  et~al., 1993, \aap, \href
  {http://adsabs.harvard.edu/abs/1993A%26A...271..482B} {271, 482}

\bibitem[\protect\citeauthoryear{{Briquet} et~al.,}{{Briquet}
  et~al.}{2012}]{2012MNRAS.427..483B}
{Briquet} M.,  et~al., 2012, \mn@doi [\mnras]
  {10.1111/j.1365-2966.2012.21933.x}, \href
  {http://adsabs.harvard.edu/abs/2012MNRAS.427..483B} {427, 483}

\bibitem[\protect\citeauthoryear{{Burssens}, {Bowman}, {Aerts}, {Pedersen},
  {Moravveji}  \& {Buysschaert}}{{Burssens} et~al.}{2019}]{2019MNRAS.489.1304B}
{Burssens} S.,  {Bowman} D.~M.,  {Aerts} C.,  {Pedersen} M.~G.,  {Moravveji}
  E.,   {Buysschaert} B.,  2019, \mn@doi [\mnras] {10.1093/mnras/stz2165},
  \href {https://ui.adsabs.harvard.edu/abs/2019MNRAS.489.1304B} {489, 1304}

\bibitem[\protect\citeauthoryear{{Burssens} et~al.,}{{Burssens}
  et~al.}{2020}]{2020A&A...639A..81B}
{Burssens} S.,  et~al., 2020, \mn@doi [\aap] {10.1051/0004-6361/202037700},
  \href {https://ui.adsabs.harvard.edu/abs/2020A&A...639A..81B} {639, A81}

\bibitem[\protect\citeauthoryear{{Buysschaert}, {Neiner}, {Martin}, {Aerts},
  {Bowman}, {Oksala}  \& {Van{\^A} Reeth}}{{Buysschaert}
  et~al.}{2018}]{2018MNRAS.478.2777B}
{Buysschaert} B.,  {Neiner} C.,  {Martin} A.~J.,  {Aerts} C.,  {Bowman} D.~M.,
  {Oksala} M.~E.,   {Van{\^A} Reeth} T.,  2018, \mn@doi [\mnras]
  {10.1093/mnras/sty1190}, \href
  {https://ui.adsabs.harvard.edu/abs/2018MNRAS.478.2777B} {478, 2777}

\bibitem[\protect\citeauthoryear{{Buysschaert}, {Neiner}, {Martin}, {Oksala},
  {Aerts}, {Tkachenko}, {Alecian}  \& {MiMeS Collaboration}}{{Buysschaert}
  et~al.}{2019}]{2019A&A...622A..67B}
{Buysschaert} B.,  {Neiner} C.,  {Martin} A.~J.,  {Oksala} M.~E.,  {Aerts} C.,
  {Tkachenko} A.,  {Alecian} E.,   {MiMeS Collaboration} 2019, \mn@doi [\aap]
  {10.1051/0004-6361/201731913}, \href
  {https://ui.adsabs.harvard.edu/abs/2019A&A...622A..67B} {622, A67}

\bibitem[\protect\citeauthoryear{{Castro} et~al.,}{{Castro}
  et~al.}{2017}]{2017A&A...597L...6C}
{Castro} N.,  et~al., 2017, \mn@doi [\aap] {10.1051/0004-6361/201629751}, \href
  {http://adsabs.harvard.edu/abs/2017A%26A...597L...6C} {597, L6}

\bibitem[\protect\citeauthoryear{{David-Uraz} et~al.,}{{David-Uraz}
  et~al.}{2019}]{2019MNRAS.487..304D}
{David-Uraz} A.,  et~al., 2019, \mn@doi [\mnras] {10.1093/mnras/stz1181}, \href
  {https://ui.adsabs.harvard.edu/abs/2019MNRAS.487..304D} {487, 304}

\bibitem[\protect\citeauthoryear{{Donati}, {Semel}  \& {Rees}}{{Donati}
  et~al.}{1992}]{dsr1992}
{Donati} J.-F.,  {Semel} M.,   {Rees} D.~E.,  1992, \aap, \href
  {http://adsabs.harvard.edu/abs/1992A26A...265..669D} {265, 669}

\bibitem[\protect\citeauthoryear{{Donati}, {Semel}, {Carter}, {Rees}  \&
  {Collier Cameron}}{{Donati} et~al.}{1997}]{d1997}
{Donati} J.-F.,  {Semel} M.,  {Carter} B.~D.,  {Rees} D.~E.,   {Collier
  Cameron} A.,  1997, MNRAS, \href
  {http://adsabs.harvard.edu/abs/1997MNRAS.291..658D} {291, 658}

\bibitem[\protect\citeauthoryear{{Donati} et~al.,}{{Donati}
  et~al.}{2008}]{2008MNRAS.390..545D}
{Donati} J.-F.,  et~al., 2008, \mn@doi [\mnras]
  {10.1111/j.1365-2966.2008.13799.x}, \href
  {http://adsabs.harvard.edu/abs/2008MNRAS.390..545D} {390, 545}

\bibitem[\protect\citeauthoryear{{Duez}, {Braithwaite}  \& {Mathis}}{{Duez}
  et~al.}{2010}]{2010ApJ...724L..34D}
{Duez} V.,  {Braithwaite} J.,   {Mathis} S.,  2010, \mn@doi [\apjl]
  {10.1088/2041-8205/724/1/L34}, \href
  {http://cdsads.u-strasbg.fr/abs/2010ApJ...724L..34D} {724, L34}

\bibitem[\protect\citeauthoryear{{Folsom} et~al.,}{{Folsom}
  et~al.}{2016}]{2016MNRAS.457..580F}
{Folsom} C.~P.,  et~al., 2016, \mn@doi [\mnras] {10.1093/mnras/stv2924}, \href
  {http://adsabs.harvard.edu/abs/2016MNRAS.457..580F} {457, 580}

\bibitem[\protect\citeauthoryear{{Folsom} et~al.,}{{Folsom}
  et~al.}{2018}]{2018MNRAS.474.4956F}
{Folsom} C.~P.,  et~al., 2018, \mn@doi [\mnras] {10.1093/mnras/stx3021}, \href
  {https://ui.adsabs.harvard.edu/abs/2018MNRAS.474.4956F} {474, 4956}

\bibitem[\protect\citeauthoryear{{Fossati} et~al.,}{{Fossati}
  et~al.}{2016}]{2016A&A...592A..84F}
{Fossati} L.,  et~al., 2016, \mn@doi [\aap] {10.1051/0004-6361/201628259},
  \href {http://adsabs.harvard.edu/abs/2016A%26A...592A..84F} {592, A84}

\bibitem[\protect\citeauthoryear{{Grunhut} et~al.,}{{Grunhut}
  et~al.}{2017}]{2017MNRAS.465.2432G}
{Grunhut} J.~H.,  et~al., 2017, \mn@doi [\mnras] {10.1093/mnras/stw2743}, \href
  {http://adsabs.harvard.edu/abs/2017MNRAS.465.2432G} {465, 2432}

\bibitem[\protect\citeauthoryear{{Hartkopf}, {Mason}, {Barry}, {McAlister},
  {Bagnuolo}  \& {Prieto}}{{Hartkopf} et~al.}{1993}]{1993AJ....106..352H}
{Hartkopf} W.~I.,  {Mason} B.~D.,  {Barry} D.~J.,  {McAlister} H.~A.,
  {Bagnuolo} W.~G.,   {Prieto} C.~M.,  1993, \mn@doi [\aj] {10.1086/116644},
  \href {https://ui.adsabs.harvard.edu/abs/1993AJ....106..352H} {106, 352}

\bibitem[\protect\citeauthoryear{{Kaufer} \& {Pasquini}}{{Kaufer} \&
  {Pasquini}}{1998}]{1998SPIE.3355..844K}
{Kaufer} A.,  {Pasquini} L.,  1998, in {D'Odorico} S.,  ed.,  \procspie Vol.
  3355, Optical Astronomical Instrumentation. pp 844--854

\bibitem[\protect\citeauthoryear{{Keszthelyi}, {Meynet}, {Georgy}, {Wade},
  {Petit}  \& {David-Uraz}}{{Keszthelyi} et~al.}{2019}]{2019MNRAS.485.5843K}
{Keszthelyi} Z.,  {Meynet} G.,  {Georgy} C.,  {Wade} G.~A.,  {Petit} V.,
  {David-Uraz} A.,  2019, \mn@doi [\mnras] {10.1093/mnras/stz772}, \href
  {http://adsabs.harvard.edu/abs/2019MNRAS.485.5843K} {485, 5843}

\bibitem[\protect\citeauthoryear{{Keszthelyi} et~al.,}{{Keszthelyi}
  et~al.}{2020}]{2020MNRAS.493..518K}
{Keszthelyi} Z.,  et~al., 2020, \mn@doi [\mnras] {10.1093/mnras/staa237}, \href
  {https://ui.adsabs.harvard.edu/abs/2020MNRAS.493..518K} {493, 518}

\bibitem[\protect\citeauthoryear{{Kharchenko}, {Piskunov}, {R{\"o}ser},
  {Schilbach}  \& {Scholz}}{{Kharchenko} et~al.}{2005}]{2005AA...438.1163K}
{Kharchenko} N.~V.,  {Piskunov} A.~E.,  {R{\"o}ser} S.,  {Schilbach} E.,
  {Scholz} R.-D.,  2005, \mn@doi [\aap] {10.1051/0004-6361:20042523}, \href
  {http://adsabs.harvard.edu/abs/2005A26A...438.1163K} {438, 1163}

\bibitem[\protect\citeauthoryear{{Kochukhov}, {Makaganiuk}  \&
  {Piskunov}}{{Kochukhov} et~al.}{2010}]{koch2010}
{Kochukhov} O.,  {Makaganiuk} V.,   {Piskunov} N.,  2010, \mn@doi [\aap]
  {10.1051/0004-6361/201015429}, \href
  {http://cdsads.u-strasbg.fr/abs/2010A26A...524A...5K} {524, A5}

\bibitem[\protect\citeauthoryear{{Kochukhov}, {Shultz}  \&
  {Neiner}}{{Kochukhov} et~al.}{2019}]{2019A&A...621A..47K}
{Kochukhov} O.,  {Shultz} M.,   {Neiner} C.,  2019, \mn@doi [\aap]
  {10.1051/0004-6361/201834279}, \href
  {https://ui.adsabs.harvard.edu/abs/2019A&A...621A..47K} {621, A47}

\bibitem[\protect\citeauthoryear{{Kupka}, {Piskunov}, {Ryabchikova}, {Stempels}
   \& {Weiss}}{{Kupka} et~al.}{1999}]{kupka1999}
{Kupka} F.~G.,  {Piskunov} N.,  {Ryabchikova} T.~A.,  {Stempels} H.~C.,
  {Weiss} W.~W.,  1999, \mn@doi [\aaps] {10.1051/aas:1999267}, \href
  {http://adsabs.harvard.edu/abs/1999A26AS..138..119K} {138, 119}

\bibitem[\protect\citeauthoryear{{Kupka}, {Ryabchikova}, {Piskunov}, {Stempels}
   \& {Weiss}}{{Kupka} et~al.}{2000}]{kupka2000}
{Kupka} F.~G.,  {Ryabchikova} T.~A.,  {Piskunov} N.~E.,  {Stempels} H.~C.,
  {Weiss} W.~W.,  2000, Baltic Astronomy, \href
  {http://adsabs.harvard.edu/abs/2000BaltA...9..590K} {9, 590}

\bibitem[\protect\citeauthoryear{{Kurapati} et~al.,}{{Kurapati}
  et~al.}{2017}]{2017MNRAS.465.2160K}
{Kurapati} S.,  et~al., 2017, \mn@doi [\mnras] {10.1093/mnras/stw2838}, \href
  {https://ui.adsabs.harvard.edu/abs/2017MNRAS.465.2160K} {465, 2160}

\bibitem[\protect\citeauthoryear{{Kuschnig}, {Weiss}, {Gruber}, {Bely}  \&
  {Jenkner}}{{Kuschnig} et~al.}{1997}]{1997A&A...328..544K}
{Kuschnig} R.,  {Weiss} W.~W.,  {Gruber} R.,  {Bely} P.~Y.,   {Jenkner} H.,
  1997, \aap, \href {http://adsabs.harvard.edu/abs/1997A%26A...328..544K} {328,
  544}

\bibitem[\protect\citeauthoryear{{Labadie-Bartz} et~al.,}{{Labadie-Bartz}
  et~al.}{2020}]{2020AJ....160...32L}
{Labadie-Bartz} J.,  et~al., 2020, \mn@doi [\aj] {10.3847/1538-3881/ab952c},
  \href {https://ui.adsabs.harvard.edu/abs/2020AJ....160...32L} {160, 32}

\bibitem[\protect\citeauthoryear{{Landstreet} \& {Borra}}{{Landstreet} \&
  {Borra}}{1978}]{lb1978}
{Landstreet} J.~D.,  {Borra} E.~F.,  1978, \mn@doi [\apjl] {10.1086/182746},
  \href {http://cdsads.u-strasbg.fr/abs/1978ApJ...224L...5L} {224, L5}

\bibitem[\protect\citeauthoryear{{Landstreet}, {Bagnulo}, {Andretta},
  {Fossati}, {Mason}, {Silaj}  \& {Wade}}{{Landstreet} et~al.}{2007}]{land2007}
{Landstreet} J.~D.,  {Bagnulo} S.,  {Andretta} V.,  {Fossati} L.,  {Mason} E.,
  {Silaj} J.,   {Wade} G.~A.,  2007, \mn@doi [\aap]
  {10.1051/0004-6361:20077343}, \href
  {http://adsabs.harvard.edu/abs/2007A26A...470..685L} {470, 685}

\bibitem[\protect\citeauthoryear{{Landstreet} et~al.,}{{Landstreet}
  et~al.}{2008}]{land2008}
{Landstreet} J.~D.,  et~al., 2008, \mn@doi [\aap] {10.1051/0004-6361:20078884},
  \href {http://adsabs.harvard.edu/abs/2008A26A...481..465L} {481, 465}

\bibitem[\protect\citeauthoryear{{Lanz} \& {Hubeny}}{{Lanz} \&
  {Hubeny}}{2007}]{lanzhubeny2007}
{Lanz} T.,  {Hubeny} I.,  2007, \mn@doi [\apjs] {10.1086/511270}, \href
  {http://adsabs.harvard.edu/abs/2007ApJS..169...83L} {169, 83}

\bibitem[\protect\citeauthoryear{{Lenz} \& {Breger}}{{Lenz} \&
  {Breger}}{2005}]{2005CoAst.146...53L}
{Lenz} P.,  {Breger} M.,  2005, \mn@doi [Communications in Asteroseismology]
  {10.1553/cia146s53}, \href
  {http://adsabs.harvard.edu/abs/2005CoAst.146...53L} {146, 53}

\bibitem[\protect\citeauthoryear{{Mathys}}{{Mathys}}{1989}]{mat1989}
{Mathys} G.,  1989, FCPh, \href
  {http://adsabs.harvard.edu/abs/1989FCPh...13..143M} {13, 143}

\bibitem[\protect\citeauthoryear{{Naz{\'e}}, {Petit}, {Rinbrand}, {Cohen},
  {Owocki}, {ud-Doula}  \& {Wade}}{{Naz{\'e}}
  et~al.}{2014}]{2014ApJS..215...10N}
{Naz{\'e}} Y.,  {Petit} V.,  {Rinbrand} M.,  {Cohen} D.,  {Owocki} S.,
  {ud-Doula} A.,   {Wade} G.~A.,  2014, \mn@doi [\apjs]
  {10.1088/0067-0049/215/1/10}, \href
  {http://adsabs.harvard.edu/abs/2014ApJS..215...10N} {215, 10}

\bibitem[\protect\citeauthoryear{{Neiner}, {Alecian}, {Briquet}, {Floquet},
  {Fr{\'e}mat}, {Martayan}, {Thizy}  \& {Mimes Collaboration}}{{Neiner}
  et~al.}{2012a}]{neiner2012b}
{Neiner} C.,  {Alecian} E.,  {Briquet} M.,  {Floquet} M.,  {Fr{\'e}mat} Y.,
  {Martayan} C.,  {Thizy} O.,   {Mimes Collaboration} 2012a, \mn@doi [\aap]
  {10.1051/0004-6361/201117941}, \href
  {http://cdsads.u-strasbg.fr/abs/2012A26A...537A.148N} {537, A148}

\bibitem[\protect\citeauthoryear{{Neiner}, {Landstreet}, {Alecian}, {Owocki},
  {Kochukhov}, {Bohlender}  \& {MiMeS collaboration}}{{Neiner}
  et~al.}{2012b}]{neiner2012a}
{Neiner} C.,  {Landstreet} J.~D.,  {Alecian} E.,  {Owocki} S.,  {Kochukhov} O.,
   {Bohlender} D.,   {MiMeS collaboration} 2012b, \mn@doi [\aap]
  {10.1051/0004-6361/201218988}, \href
  {http://cdsads.u-strasbg.fr/abs/2012A26A...546A..44N} {546, A44}

\bibitem[\protect\citeauthoryear{{Neiner}, {Mathis}, {Alecian}, {Emeriau},
  {Grunhut}, {BinaMIcS}  \& {MiMeS Collaborations}}{{Neiner}
  et~al.}{2015}]{2015IAUS..305...61N}
{Neiner} C.,  {Mathis} S.,  {Alecian} E.,  {Emeriau} C.,  {Grunhut} J.,
  {BinaMIcS}  {MiMeS Collaborations} 2015, in {Nagendra} K.~N.,  {Bagnulo} S.,
  {Centeno} R.,   {Jes{\'u}s Mart{\'{\i}}nez Gonz{\'a}lez} M.,  eds,  IAU
  Symposium Vol. 305, Polarimetry. pp 61--66 (\mn@eprint {arXiv} {1502.00226}),
  \mn@doi{10.1017/S1743921315004524}

\bibitem[\protect\citeauthoryear{{Owocki}, {Shultz}, {ud-Doula}, {Sundqvist},
  {Townsend}  \& {Cranmer}}{{Owocki} et~al.}{2020}]{2020arXiv200912359O}
{Owocki} S.~P.,  {Shultz} M.~E.,  {ud-Doula} A.,  {Sundqvist} J.~O.,
  {Townsend} R. H.~D.,   {Cranmer} S.~R.,  2020, arXiv e-prints, \href
  {https://ui.adsabs.harvard.edu/abs/2020arXiv200912359O} {p. arXiv:2009.12359}

\bibitem[\protect\citeauthoryear{{Pablo} et~al.,}{{Pablo}
  et~al.}{2019}]{2019MNRAS.488...64P}
{Pablo} H.,  et~al., 2019, \mn@doi [\mnras] {10.1093/mnras/stz1661}, \href
  {https://ui.adsabs.harvard.edu/abs/2019MNRAS.488...64P} {488, 64}

\bibitem[\protect\citeauthoryear{{Pedersen} et~al.,}{{Pedersen}
  et~al.}{2019}]{2019ApJ...872L...9P}
{Pedersen} M.~G.,  et~al., 2019, \mn@doi [\apjl] {10.3847/2041-8213/ab01e1},
  \href {https://ui.adsabs.harvard.edu/abs/2019ApJ...872L...9P} {872, L9}

\bibitem[\protect\citeauthoryear{{Petit} et~al.,}{{Petit}
  et~al.}{2013}]{petit2013}
{Petit} V.,  et~al., 2013, \mn@doi [\mnras] {10.1093/mnras/sts344}, \href
  {http://adsabs.harvard.edu/abs/2013MNRAS.429..398P} {429, 398}

\bibitem[\protect\citeauthoryear{{Pigulski}}{{Pigulski}}{1992}]{1992A&A...261..203P}
{Pigulski} A.,  1992, \aap, \href
  {https://ui.adsabs.harvard.edu/abs/1992A&A...261..203P} {261, 203}

\bibitem[\protect\citeauthoryear{{Piskunov}, {Kupka}, {Ryabchikova}, {Weiss}
  \& {Jeffery}}{{Piskunov} et~al.}{1995}]{piskunov1995}
{Piskunov} N.~E.,  {Kupka} F.,  {Ryabchikova} T.~A.,  {Weiss} W.~W.,
  {Jeffery} C.~S.,  1995, \aaps, \href
  {http://adsabs.harvard.edu/abs/1995A26AS..112..525P} {112, 525}

\bibitem[\protect\citeauthoryear{{Preston}}{{Preston}}{1967}]{preston1967}
{Preston} G.~W.,  1967, \mn@doi [\apj] {10.1086/149358}, \href
  {http://adsabs.harvard.edu/abs/1967ApJ...150..547P} {150, 547}

\bibitem[\protect\citeauthoryear{{Preston}}{{Preston}}{1974}]{preston1974}
{Preston} G.~W.,  1974, \mn@doi [\araa] {10.1146/annurev.aa.12.090174.001353},
  \href {http://cdsads.u-strasbg.fr/abs/1974ARA26A..12..257P} {12, 257}

\bibitem[\protect\citeauthoryear{{Ricker} et~al.,}{{Ricker}
  et~al.}{2015}]{2015JATIS...1a4003R}
{Ricker} G.~R.,  et~al., 2015, \mn@doi [Journal of Astronomical Telescopes,
  Instruments, and Systems] {10.1117/1.JATIS.1.1.014003}, \href
  {https://ui.adsabs.harvard.edu/abs/2015JATIS...1a4003R} {1, 014003}

\bibitem[\protect\citeauthoryear{{Rivinius}, {Baade}, {Hadrava}, {Heida}  \&
  {Klement}}{{Rivinius} et~al.}{2020}]{2020A&A...637L...3R}
{Rivinius} T.,  {Baade} D.,  {Hadrava} P.,  {Heida} M.,   {Klement} R.,  2020,
  \mn@doi [\aap] {10.1051/0004-6361/202038020}, \href
  {https://ui.adsabs.harvard.edu/abs/2020A&A...637L...3R} {637, L3}

\bibitem[\protect\citeauthoryear{{Ryabchikova}, {Piskunov}, {Kupka}  \&
  {Weiss}}{{Ryabchikova} et~al.}{1997}]{ryabchikova1997}
{Ryabchikova} T.~A.,  {Piskunov} N.~E.,  {Kupka} F.,   {Weiss} W.~W.,  1997,
  Baltic Astronomy, \href {http://cdsads.u-strasbg.fr/abs/1997BaltA...6..244R}
  {6, 244}

\bibitem[\protect\citeauthoryear{{Ryabchikova}, {Piskunov}, {Kurucz},
  {Stempels}, {Heiter}, {Pakhomov}  \& {Barklem}}{{Ryabchikova}
  et~al.}{2015}]{2015PhyS...90e4005R}
{Ryabchikova} T.,  {Piskunov} N.,  {Kurucz} R.~L.,  {Stempels} H.~C.,  {Heiter}
  U.,  {Pakhomov} Y.,   {Barklem} P.~S.,  2015, \mn@doi [\physscr]
  {10.1088/0031-8949/90/5/054005}, \href
  {https://ui.adsabs.harvard.edu/abs/2015PhyS...90e4005R} {90, 054005}

\bibitem[\protect\citeauthoryear{{Shultz} et~al.,}{{Shultz}
  et~al.}{2015a}]{2015MNRAS.449.3945S}
{Shultz} M.,  et~al., 2015a, \mn@doi [\mnras] {10.1093/mnras/stv564}, \href
  {http://adsabs.harvard.edu/abs/2015MNRAS.449.3945S} {449, 3945}

\bibitem[\protect\citeauthoryear{{Shultz}, {Wade}, {Alecian}  \& {BinaMIcS
  Collaboration}}{{Shultz} et~al.}{2015b}]{2015MNRAS.454L...1S}
{Shultz} M.,  {Wade} G.~A.,  {Alecian} E.,   {BinaMIcS Collaboration} 2015b,
  \mn@doi [\mnras] {10.1093/mnrasl/slv096}, \href
  {http://adsabs.harvard.edu/abs/2015MNRAS.454L...1S} {454, L1}

\bibitem[\protect\citeauthoryear{{Shultz}, {Wade}, {Rivinius}, {Neiner},
  {Henrichs}, {Marcolino}  \& {MiMeS Collaboration}}{{Shultz}
  et~al.}{2017}]{2017MNRAS.471.2286S}
{Shultz} M.,  {Wade} G.~A.,  {Rivinius} T.,  {Neiner} C.,  {Henrichs} H.,
  {Marcolino} W.,   {MiMeS Collaboration} 2017, \mn@doi [\mnras]
  {10.1093/mnras/stx1632}, \href
  {http://adsabs.harvard.edu/abs/2017MNRAS.471.2286S} {471, 2286}

\bibitem[\protect\citeauthoryear{{Shultz}, {Rivinius}, {Wade}, {Alecian}  \&
  {Petit}}{{Shultz} et~al.}{2018a}]{2018MNRAS.475..839S}
{Shultz} M.,  {Rivinius} T.,  {Wade} G.~A.,  {Alecian} E.,   {Petit} V.,
  2018a, \mn@doi [\mnras] {10.1093/mnras/stx3238}, \href
  {http://adsabs.harvard.edu/abs/2018MNRAS.475..839S} {475, 839}

\bibitem[\protect\citeauthoryear{{Shultz} et~al.,}{{Shultz}
  et~al.}{2018b}]{2018MNRAS.475.5144S}
{Shultz} M.~E.,  et~al., 2018b, \mn@doi [\mnras] {10.1093/mnras/sty103}, \href
  {http://adsabs.harvard.edu/abs/2018MNRAS.475.5144S} {475, 5144}

\bibitem[\protect\citeauthoryear{{Shultz} et~al.,}{{Shultz}
  et~al.}{2019a}]{2019MNRAS.485.1508S}
{Shultz} M.~E.,  et~al., 2019a, \mn@doi [\mnras] {10.1093/mnras/stz416}, \href
  {https://ui.adsabs.harvard.edu/abs/2019MNRAS.485.1508S} {485, 1508}

\bibitem[\protect\citeauthoryear{{Shultz} et~al.,}{{Shultz}
  et~al.}{2019b}]{2019MNRAS.490..274S}
{Shultz} M.~E.,  et~al., 2019b, \mn@doi [\mnras] {10.1093/mnras/stz2551}, \href
  {https://ui.adsabs.harvard.edu/abs/2019MNRAS.490..274S} {490, 274}

\bibitem[\protect\citeauthoryear{{Shultz} et~al.,}{{Shultz}
  et~al.}{2020}]{2020arXiv200912336S}
{Shultz} M.~E.,  et~al., 2020, arXiv e-prints, \href
  {https://ui.adsabs.harvard.edu/abs/2020arXiv200912336S} {p. arXiv:2009.12336}

\bibitem[\protect\citeauthoryear{{Sikora}, {Wade}, {Power}  \&
  {Neiner}}{{Sikora} et~al.}{2019a}]{2019MNRAS.483.2300S}
{Sikora} J.,  {Wade} G.~A.,  {Power} J.,   {Neiner} C.,  2019a, \mn@doi
  [\mnras] {10.1093/mnras/sty3105}, \href
  {http://adsabs.harvard.edu/abs/2019MNRAS.483.2300S} {483, 2300}

\bibitem[\protect\citeauthoryear{{Sikora}, {Wade}, {Power}  \&
  {Neiner}}{{Sikora} et~al.}{2019b}]{2019MNRAS.483.3127S}
{Sikora} J.,  {Wade} G.~A.,  {Power} J.,   {Neiner} C.,  2019b, \mn@doi
  [\mnras] {10.1093/mnras/sty2895}, \href
  {http://adsabs.harvard.edu/abs/2019MNRAS.483.3127S} {483, 3127}

\bibitem[\protect\citeauthoryear{{Sikora} et~al.,}{{Sikora}
  et~al.}{2019c}]{2019MNRAS.487.4695S}
{Sikora} J.,  et~al., 2019c, \mn@doi [\mnras] {10.1093/mnras/stz1581}, \href
  {https://ui.adsabs.harvard.edu/abs/2019MNRAS.487.4695S} {487, 4695}

\bibitem[\protect\citeauthoryear{{Sundqvist}, {Petit}, {Owocki}, {Wade}, {Puls}
   \& {MiMeS Collaboration}}{{Sundqvist} et~al.}{2013}]{2013MNRAS.433.2497S}
{Sundqvist} J.~O.,  {Petit} V.,  {Owocki} S.~P.,  {Wade} G.~A.,  {Puls} J.,
  {MiMeS Collaboration} 2013, \mn@doi [\mnras] {10.1093/mnras/stt921}, \href
  {https://ui.adsabs.harvard.edu/abs/2013MNRAS.433.2497S} {433, 2497}

\bibitem[\protect\citeauthoryear{{Takahashi}}{{Takahashi}}{2020}]{2020svos.conf..293T}
{Takahashi} K.,  2020, in {Neiner} C.,  {Weiss} W.~W.,  {Baade} D.,  {Griffin}
  R.~E.,  {Lovekin} C.~C.,   {Moffat} A.~F.~J.,  eds, Proceedings of the
  conference Stars and their Variability Observed from Space. pp 293--296

\bibitem[\protect\citeauthoryear{{Tokovinin}, {Mason}  \&
  {Hartkopf}}{{Tokovinin} et~al.}{2010}]{2010AJ....139..743T}
{Tokovinin} A.,  {Mason} B.~D.,   {Hartkopf} W.~I.,  2010, \mn@doi [\aj]
  {10.1088/0004-6256/139/2/743}, \href
  {https://ui.adsabs.harvard.edu/abs/2010AJ....139..743T} {139, 743}

\bibitem[\protect\citeauthoryear{{Townsend} \& {Owocki}}{{Townsend} \&
  {Owocki}}{2005}]{town2005c}
{Townsend} R.~H.~D.,  {Owocki} S.~P.,  2005, \mn@doi [\mnras]
  {10.1111/j.1365-2966.2005.08642.x}, \href
  {http://adsabs.harvard.edu/abs/2005MNRAS.357..251T} {357, 251}

\bibitem[\protect\citeauthoryear{{Vink}, {de Koter}  \& {Lamers}}{{Vink}
  et~al.}{2001}]{vink2001}
{Vink} J.~S.,  {de Koter} A.,   {Lamers} H.~J.~G.~L.~M.,  2001, \mn@doi [\aap]
  {10.1051/0004-6361:20010127}, \href
  {http://adsabs.harvard.edu/abs/2001A26A...369..574V} {369, 574}

\bibitem[\protect\citeauthoryear{{Wade} et~al.,}{{Wade}
  et~al.}{2016}]{2016MNRAS.456....2W}
{Wade} G.~A.,  et~al., 2016, \mn@doi [\mnras] {10.1093/mnras/stv2568}, \href
  {http://adsabs.harvard.edu/abs/2016MNRAS.456....2W} {456, 2}

\bibitem[\protect\citeauthoryear{{ud-Doula} \& {Owocki}}{{ud-Doula} \&
  {Owocki}}{2002}]{ud2002}
{ud-Doula} A.,  {Owocki} S.~P.,  2002, \mn@doi [ApJ] {10.1086/341543}, \href
  {http://adsabs.harvard.edu/abs/2002ApJ...576..413U} {576, 413}

\bibitem[\protect\citeauthoryear{{ud-Doula}, {Owocki}  \&
  {Townsend}}{{ud-Doula} et~al.}{2008}]{ud2008}
{ud-Doula} A.,  {Owocki} S.~P.,   {Townsend} R.~H.~D.,  2008, \mn@doi [MNRAS]
  {10.1111/j.1365-2966.2008.12840.x}, \href
  {http://adsabs.harvard.edu/abs/2008MNRAS.385...97U} {385, 97}

\makeatother
\end{thebibliography}


\appendix

\section{Radial Velocities}

\begin{table}
\centering
\caption[]{Radial Velocity (RV) measurements. }
\label{rvtab}
\begin{tabular}{l r r}
\hline\hline
Instrument & HJD$-$ & RV \\
           & 2456000 & (\kms) \\
\hline
 HARPSpol &  126.64801 &   1.6$\pm$  0.4 \\
 HARPSpol &  126.65882 &  -0.8$\pm$  0.7 \\
 HARPSpol &  126.66962 &  -2.7$\pm$  0.4 \\
 HARPSpol &  126.68042 &  -2.9$\pm$  0.3 \\
 HARPSpol &  126.69123 &  -1.0$\pm$  0.4 \\
 HARPSpol &  126.70204 &   1.2$\pm$  0.7 \\
 HARPSpol &  126.71284 &   3.5$\pm$  1.1 \\
 HARPSpol &  126.72363 &   3.8$\pm$  0.7 \\
 HARPSpol &  127.76099 &  -1.2$\pm$  1.1 \\
 HARPSpol &  127.77180 &   0.7$\pm$  0.3 \\
 HARPSpol &  127.78261 &   3.3$\pm$  0.3 \\
 HARPSpol &  127.79342 &   4.0$\pm$  0.8 \\
 ESPaDOnS &  758.03027 &  12.3$\pm$  0.8 \\
 ESPaDOnS &  758.03596 &  11.5$\pm$  1.1 \\
 ESPaDOnS &  758.04166 &  10.3$\pm$  1.0 \\
 ESPaDOnS &  758.04734 &   9.4$\pm$  1.1 \\
 ESPaDOnS &  761.94523 &   9.1$\pm$  1.4 \\
 ESPaDOnS &  761.95091 &  10.7$\pm$  1.6 \\
 ESPaDOnS &  761.95661 &  11.3$\pm$  0.7 \\
 ESPaDOnS &  761.96230 &  10.2$\pm$  0.7 \\
 ESPaDOnS &  761.96893 &   9.9$\pm$  2.0 \\
 ESPaDOnS &  761.97461 &  10.1$\pm$  0.8 \\
 ESPaDOnS &  761.98029 &   7.0$\pm$  0.8 \\
 ESPaDOnS &  761.98598 &   5.8$\pm$  1.0 \\
 ESPaDOnS &  814.95607 &  10.6$\pm$  1.2 \\
 ESPaDOnS &  814.96176 &  11.4$\pm$  0.8 \\
 ESPaDOnS &  814.96745 &  10.4$\pm$  0.8 \\
 ESPaDOnS &  814.97313 &  10.8$\pm$  0.8 \\
 ESPaDOnS &  814.97908 &   9.0$\pm$  1.0 \\
 ESPaDOnS &  814.98477 &   7.6$\pm$  0.9 \\
 ESPaDOnS &  814.99046 &   5.7$\pm$  1.0 \\
 ESPaDOnS &  814.99615 &   4.3$\pm$  1.1 \\
 ESPaDOnS &  821.93185 &  10.0$\pm$  0.9 \\
 ESPaDOnS &  821.93754 &   9.5$\pm$  1.1 \\
 ESPaDOnS &  821.94323 &   7.8$\pm$  1.3 \\
 ESPaDOnS &  821.94892 &   6.3$\pm$  1.1 \\
 ESPaDOnS &  821.97258 &   4.5$\pm$  1.1 \\
 ESPaDOnS &  821.97827 &   4.8$\pm$  1.2 \\
 ESPaDOnS &  821.98395 &   5.8$\pm$  1.0 \\
 ESPaDOnS &  821.98964 &   7.4$\pm$  0.9 \\
 ESPaDOnS &  824.88691 &   8.7$\pm$  0.6 \\
 ESPaDOnS &  824.89260 &   7.2$\pm$  0.7 \\
 ESPaDOnS &  824.89829 &   4.8$\pm$  1.2 \\
 ESPaDOnS &  824.90398 &   3.6$\pm$  1.1 \\
 ESPaDOnS &  824.90983 &   3.0$\pm$  1.4 \\
 ESPaDOnS &  824.91552 &   3.3$\pm$  0.6 \\
 ESPaDOnS &  824.92121 &   3.8$\pm$  0.9 \\
 ESPaDOnS &  824.92690 &   5.2$\pm$  0.9 \\
 FEROS & 1203.63038 &   8.9$\pm$  1.0 \\
 FEROS & 1203.69126 &   2.9$\pm$  0.8 \\
 FEROS & 1204.67172 &   3.5$\pm$  0.9 \\
 FEROS & 1204.73291 &   7.7$\pm$  0.9 \\
 FEROS & 1205.49821 &   6.5$\pm$  1.0 \\
 FEROS & 1205.66785 &   4.5$\pm$  0.5 \\
 FEROS & 1205.69098 &   9.5$\pm$  0.6 \\
 FEROS & 1206.50130 &  10.2$\pm$  1.3 \\
 FEROS & 1206.70782 &   4.7$\pm$  1.0 \\
 FEROS & 1207.55325 &   8.6$\pm$  1.3 \\
 FEROS & 1207.71807 &   6.0$\pm$  0.9 \\
 FEROS & 1208.55215 &   9.2$\pm$  1.1 \\
\hline\hline
\end{tabular}
\end{table}

\end{document}